\documentclass[journal,comsoc]{IEEEtran}

\usepackage[english]{babel}
\usepackage[latin1]{inputenc}
\usepackage{bbm}
\usepackage{url}
\usepackage{amsmath,pifont,amssymb,amsfonts}
\usepackage{graphicx}
\usepackage{color}
\usepackage{framed}
\usepackage{gensymb}
\usepackage{textcomp}
\usepackage{longtable}
\usepackage{array}
\usepackage{subcaption}
\usepackage{psfrag}
\usepackage{algorithm}
\usepackage{algorithmic}
\usepackage{cite}
\usepackage{multirow}
\usepackage{tabularx}
\usepackage{enumerate}

\usepackage{pdfpages} % add full-page images support
\usepackage{lscape} % add landscape page rotation

\usepackage{color}

\usepackage{xspace}

\usepackage[printonlyused]{acronym}
\acrodef{NTMA}{Network Traffic Monitoring and Analysis}
\acrodef{RDBMS}{Relational Database Management Systems}
\acrodef{KDD}{Knowledge Discovery in Data}
\acrodef{CDN}{Content Delivery Network}

\usepackage{array}
\newcolumntype{x}[1]{>{\centering\let\newline\\\arraybackslash\hspace{0pt}}m{#1}}

\graphicspath{{figs/}}

\usepackage{flushend}

\begin{document}

\title{A Survey on Big Data for Network Traffic Monitoring and Analysis\thanks{The research leading to these results has been funded by the Vienna Science and Technology Fund (WWTF) through project ICT15-129 ``BigDAMA'', with the contribution from the SmartData@PoliTO center for data science and big data. Authors sincerely thank Prof. Tanja Zseby, Dr. Felix Iglesias, and Daniel C. Ferreira from TU Wien for their contributions to the work and early versions of the manuscript.}}

\author{
Alessandro D'Alconzo\IEEEauthorrefmark{3},
Idilio Drago\IEEEauthorrefmark{2},
Andrea Morichetta\IEEEauthorrefmark{2},
Marco Mellia\IEEEauthorrefmark{2},
Pedro Casas\IEEEauthorrefmark{1}\\
\IEEEauthorrefmark{1} AIT Austrian Institute of Technology, \IEEEauthorrefmark{2} Politecnico di Torino, \IEEEauthorrefmark{3} Siemens Austria
}

\maketitle

\begin{abstract}
\ac{NTMA} represents a key component for network management, especially to guarantee the correct operation of large-scale networks such as the Internet. As the complexity of Internet services and the volume of traffic continue to increase, it becomes difficult to design scalable \ac{NTMA} applications. Applications such as traffic classification and policing require real-time and scalable approaches. Anomaly detection and security mechanisms require to quickly identify and react to unpredictable events while processing millions of heterogeneous events. At last, the system has to collect, store, and process massive sets of historical data for post-mortem analysis. Those are precisely the challenges faced by general \emph{big data} approaches: Volume, Velocity, Variety, and Veracity. This survey brings together \ac{NTMA} and big data. We catalog previous work on \ac{NTMA} that adopt big data approaches to understand to what extent the potential of big data is being explored in \ac{NTMA}. This survey mainly focuses on approaches and technologies to manage the big \ac{NTMA} data, additionally briefly discussing big data analytics (e.g., machine learning) for the sake of NTMA. Finally, we provide guidelines for future work, discussing lessons learned, and research directions.
\end{abstract}


\section{Introduction}

Understanding how Internet services are used and how they are operating is critical to people lives. Network Traffic Monitoring and Analysis (\ac{NTMA}) is central to that task. Applications range from providing a view on network traffic to the detection of anomalies and unknown attacks while feeding systems responsible for usage monitoring and accounting. They collect the historical data needed to support traffic engineering and troubleshooting, helping to plan the network evolution and identify the root cause of problems. It is correct to say that \ac{NTMA} applications are a cornerstone to guarantee that the services supporting our daily lives are always available and operating as expected.

Traffic monitoring and analysis is a complicated task. The massive traffic volumes, the speed of transmission systems, the natural evolution of services and attacks, and the variety of data sources and methods to acquire measurements are just some of the challenges faced by \ac{NTMA} applications. As the complexity of the network continues to increase, more observation points become available to researchers, potentially allowing heterogeneous data to be collected and evaluated. This trend makes it hard to design scalable and distributed applications and calls for efficient mechanisms for online analysis of large streams of measurements. More than that, as storage prices decrease, it becomes possible to create massive historical datasets for retrospective analysis.

These challenges are precisely the characteristics associated with what, more recently, have become known as \emph{big data}, i.e., situations in which the data \emph{volume}, \emph{velocity}, \emph{veracity} and \emph{variety} are the key challenges to allow the extraction of \emph{value} from the data. Indeed, traffic monitoring and analysis were one of the first examples of big data sources to emerge, and it poses big data challenges more than ever.

It is thus not a surprise that researchers are resorting to big data technologies to support \ac{NTMA} applications (e.g.,~\cite{lee_toward_2013, marchal_big_2014, orsini_bgpstream_2016, wullink_entrada_2016}). Distributed file systems -- e.g., the Hadoop\footnote{\url{http://hadoop.apache.org/}} Distributed File System (HDFS), big data platforms -- e.g., Hadoop and Spark, and distributed machine learning and graph processing engines -- e.g., MLlib and Apache Giraph, are some examples of technologies that are assisting applications to handle datasets that otherwise would be intractable. However, it is by no means clear whether \ac{NTMA} applications fully exploit the potential of emerging big data technologies.

% existing big data technologies cover all the needs of \ac{NTMA} applications, as it is neither clear

We bring together \ac{NTMA} research and big data. Whereas previous works documented advances on big data research and technologies~\cite{zhang_parallel_2016, fahad_survey_2014, tsai_big_2015}, methods supporting \ac{NTMA} (e.g., machine learning for \ac{NTMA}~\cite{nguyen_survey_2008,callado_survey_2009}), or specific \ac{NTMA} applications~\cite{sperotto_overview_2010, valenti_reviewing_2013, hofstede_flow_2014, bhuyan_network_2014}, there is a lack of systematic surveys describing how \ac{NTMA} and big data are being combined to exploit the potential of network data fully.

% More concretely, the goal of this survey is therefore twofold: (i)~to present and discuss \emph{to what extent big data technologies cover the basic requirements of \ac{NTMA} applications}; (ii)~to verify \emph{whether and to what extent \ac{NTMA} applications exploit the potential of big data technologies}.

More concretely, the goal of this survey is to discuss \emph{to what extent \ac{NTMA} researchers are exploiting the potential of big data technologies}. We aim at providing network researchers willing to adopt big data approaches for \ac{NTMA} applications a broad overview of success cases and pitfalls on previous research efforts, thus illustrating the challenges and opportunities on the use of big data technologies for \ac{NTMA} applications.

By summarizing recent literature on \ac{NTMA}, we provide researchers principled guidelines on the use of big data according to the requirements and purposes of \ac{NTMA} applications. Ultimately, by cataloging how challenges on \ac{NTMA} have been faced with big data approaches, we highlight open issues and promising research directions.

\subsection{Survey methodology}

We first identify papers summarizing big data research. We have explored the literature for big data surveys, restricting our focus to the last ten years. This literature has served as a basis to our definition for big data as well as to limit our scope in terms of the considered big data technologies.

Since none of these papers addresses big data for \ac{NTMA}, we have followed a similar methodology and reviewed papers in the \ac{NTMA} domain. We survey how \ac{NTMA} researchers are profiting from big data approaches and technologies. We have focused on the last 5 years of (i) the main system and networking conferences, namely SIGCOMM, NSDI, CONEXT, and journals (IEEE TON, IEEE TNSM), (ii) new venues targeting analytics and big data for \ac{NTMA}, namely the BigDama, AnNet, WAIN and NetAI workshops, and their parent conferences (e.g., IM, NOMS and CNSM); (ii) special issues on Big Data Analytics published in this journal (TNSM). We complement the survey by searching on Google Scholar, IEEE Explorer, and ACM Digital library. To ensure relevance and limit our scope, we select papers concerning publication venues, the number of citations, and scope.

%We emphasize the use of big data technologies and frameworks, briefly reviewing big data analytics for \ac{NTMA} for the sake of completeness.

% Reviewed papers can be categorized as work approaching big data technologies and frameworks (e.g., \cite{Zhang16,Bajaber16,hu2014,lee2012,singh2014,Jing11}), and work focusing on analytics (e.g., \cite{tsai2015,Fahad14,Shirkhorshidi14}). We have reviewed both areas to get a systematic overview of these aspects. We acknowledge that both domains are large and complex, with a vast and assorted literature;

The survey is organized as follows: Sect.~\ref{sec:bigdata} introduces concepts of big data and the steps for big data analytics, giving also some background in big data platforms. Sect.~\ref{sec:intro_tma} reviews a taxonomy of \ac{NTMA} applications, illustrating how \ac{NTMA} relates to the big data challenges. Subsequent sections detail the process of \ac{NTMA} and discuss how previous work has faced big data challenges in \ac{NTMA}. Sect.~\ref{sec:big_data_for_NTMA} focuses on data capture, ingestion, storage and pre-processing, whereas Sect.~\ref{sec:big_data_analytics_for_NTMA} overviews big data analytics for \ac{NTMA}. Finally, Sect.~\ref{sec:discussion} describes lessons learned and open issues.

% Sect.~\ref{sec:big_data_technologies} summarizes big data technologies.

\section{What is big data?}\label{sec:bigdata}

\subsection{Definition}

Many definitions for big data have appeared in the literature. We rely on previous surveys that focused on other aspects of big data but have already documented its definitions.

Hu~at~al.~\cite{hu_toward_2014} argue that ``big data means not only a large volume of data but also other features, such as variety and velocity of the data''. They organize definitions in three categories: architectural, comparative, or attributive.

Architectural and comparative definitions are both abstract. Following an architectural definition, one would face a big data problem whenever the dataset is such that ``traditional approaches'' are incapable of storing and handling it. Similarly, comparative definitions characterize big data by comparing the properties of the data to the capabilities of traditional database tools.
Those relying on \emph{attributive} definitions (e.g., ~\cite{manyika_big_2011,laney_3d_2001}) describe big data by the salient features of both the data and the process of analyzing the data itself. They advocate what is nowadays widely known by the big data \emph{``V's''} -- e.g., volume, velocity, variety, veracity, value.

Other surveys targeting big data technologies~\cite{zhang_parallel_2016,bajaber_big_2016,hu_toward_2014} and analytics~\cite{tsai_big_2015,fahad_survey_2014} share this view. We will stick to the ``5-Vs'' definition because it provides concrete criteria that characterize the big data challenges. We thus consider a problem to belong to the big data class if datasets are large (i.e., volume) and need to be captured and analyzed at high rates or in real-time (i.e., velocity). The data potentially come from different sources (i.e., variety) that combine (i)~structured data such as column-oriented databases; (ii)~unstructured data such as server logs; and (iii)~semi-structured data such as XML or JSON documents. Moreover, data can be of different quality, with the uncertainty that characterizes the data (i.e., veracity). At last, the analysis of the data brings advantages for users and businesses (i.e., value). That is, new insights are extracted from the original data to increase its value, preferably using automatic methodologies.

We argue next that \ac{NTMA} shares these characteristics. Hence, \ac{NTMA} is an example of big data application which can profit from methodologies developed to face these challenges.

\subsection{Knowledge discovery in big data}

\begin{figure}[!t]
  \centering
  \includegraphics[width=0.45\linewidth]{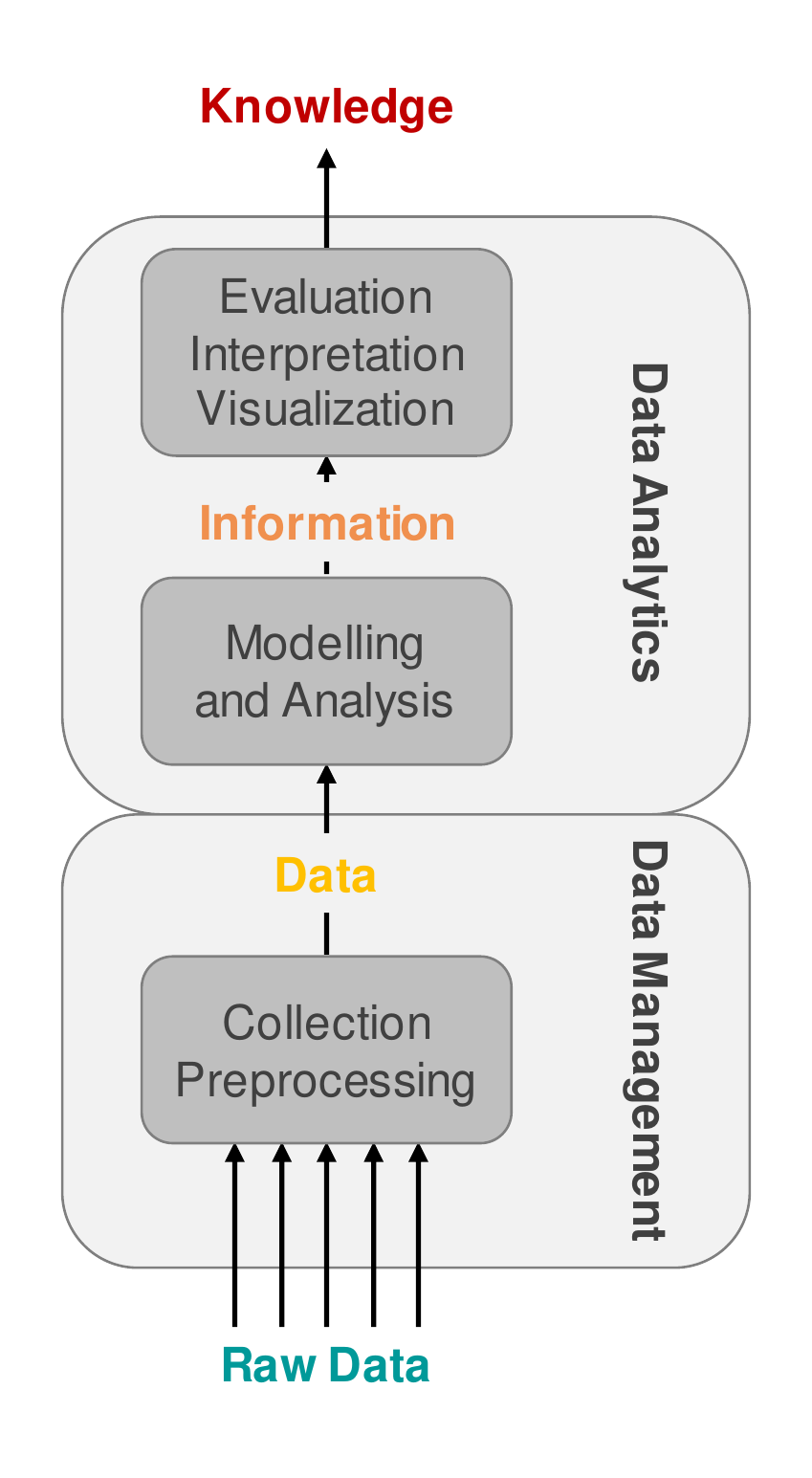}
  \caption{\ac{KDD}~\cite{fayyad_data_1996,tsai_big_2015}.}
  \label{fig:KDD_process}
\end{figure}

The process of \ac{KDD} is often attributed to Fayyad~et~al.~\cite{fayyad_data_1996} who summarized it as a sequence of operations over the data: gathering, selection, pre-processing, transformation, mining, evaluation, interpretation, and visualization. From a system perspective, authors of~\cite{tsai_big_2015} reorganized these operations in three groups: input, analysis, and output.

\textbf{Data input} performs the data management process, from the collection of raw data to the delivery of data in suitable formats for subsequent mining. It includes pre-processing, which are the initial steps to prepare the data, with the integration of heterogeneous sources and cleaning of spurious data.

\textbf{Data analysis} methods receive the prepared data and extract information, i.e., models for classification, hidden patterns, relations, rules, etc. The methods range from statistical modeling and analysis to machine learning and data mining algorithms.

\textbf{Data output} completes the process by converting information into knowledge. It includes steps to measure the information quality, to display information in succinct formats, and to assist analysts with the interpretation of results. This step is not evaluated in this survey and is ignored in the following.

The stages identified by Fayyad~et~al. can be further grouped into data management and analytics (i.e., data analysis and output). Note that differently from~\cite{tsai_big_2015}, we use the term \emph{data analysis} to refer to the algorithms for extracting information from the data, whereas we use \emph{data analytics} to refer to the whole process, from data analysis to knowledge discovery.

We reproduce and adapt this scheme in Fig.~\ref{fig:KDD_process} and will use it to characterize \ac{NTMA} papers according to the several stages of the \ac{KDD} process.

The \ac{KDD} process described above applies to any data analytics. However, the characteristics of the big data impose fundamental challenges to the methodologies on each step. For example, the data management process will naturally face much more complex tasks with big data, given the volume, velocity, and variety of the data. Data management is, therefore, crucial with big data since it plays a key role in reducing data volume and complexity. Also, it impacts the analysis and the output phases, in particular in terms of speed of the analysis and value of results.

The big data challenges (i.e., the ``5-Vs'') call for an approach that considers the whole \ac{KDD} process in a comprehensive analytics framework. Such a framework should include programming models that allow implementing application logic covering the complete \ac{KDD} cycle. We will show later that a common practice to cope with big datasets is to resort to parallel and distributed computing frameworks that are still on expansion to cover all \ac{KDD} phases.

\subsection{Programming models and platforms}\label{sec:programming_models}

\begin{figure}[!t]
    \centering
    \includegraphics[width=1.15\linewidth]{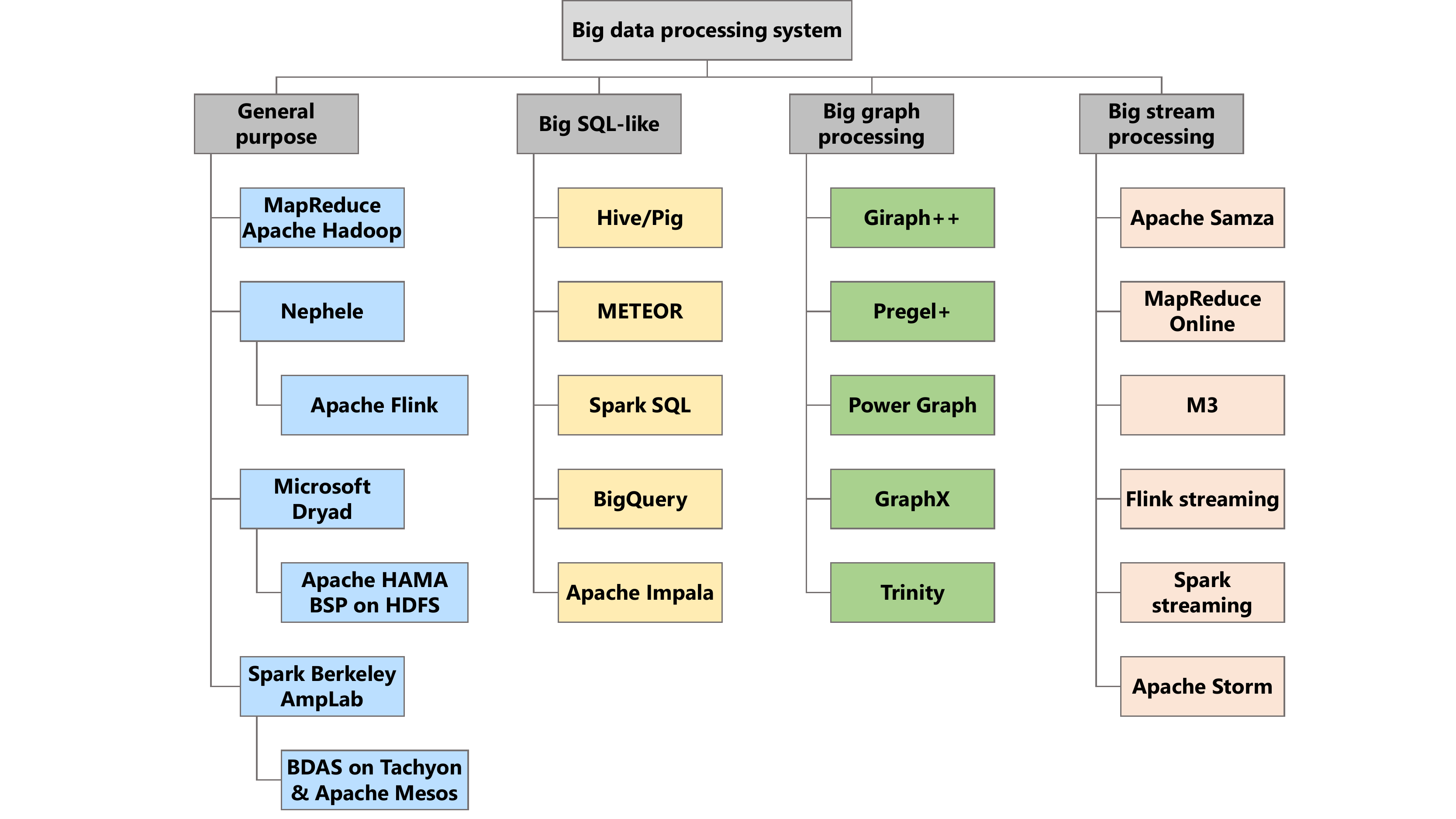}
    \caption{Processing systems (based on~\cite{bajaber_big_2016, sakr_family_2013, zhang_parallel_2016}).}
    \label{fig:compute_infrasructures}
\end{figure}

We provide a short overview of the most relevant programming models and platforms for handling big data. While surveying the literature, we will give particular emphasis on works that make use of such models and platforms for \ac{NTMA}.

Following the taxonomies found in~\cite{bajaber_big_2016, zhang_parallel_2016}, we distinguish four types of big data processing models: (i) general purpose -- platforms to process big data that make little assumptions about the data characteristics and the executed algorithms, (ii) SQL-like -- platforms focusing on scalable processing of structured and tabular data, (iii) graph processing -- platforms focusing on the processing of large graphs, and (iv) stream processing --  platforms dealing large-scale data that continuously arrive to the system in a streaming fashion.

Fig.~\ref{fig:compute_infrasructures} depicts this taxonomy with examples of systems. MapReduce, Dryad, Flink, and Spark belong to the first type. Hive, HAWQ, Apache Drill, and Tajo belong to the SQL-like type. Pregel, GraphLab follow the graph processing models and, finally, Storm and S4 are examples of the latter.

A comprehensive review of big data programming models and platforms is far beyond the scope of this paper. We refer readers to~\cite{bajaber_big_2016, zhang_parallel_2016} for a complete survey.

\subsubsection{The Hadoop ecosystem}

\begin{figure}
    \centering
    \includegraphics[width=.8\linewidth]{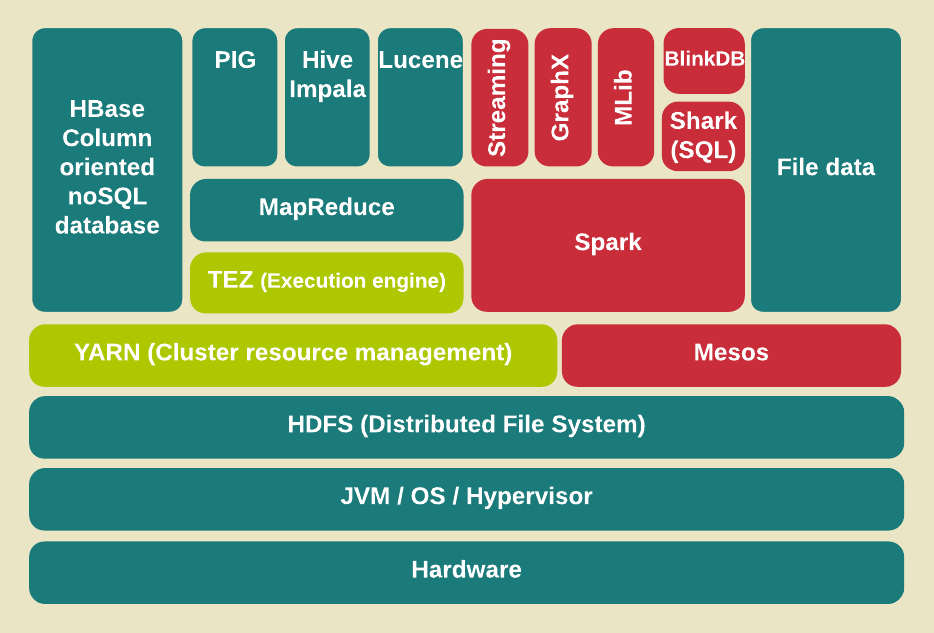}
    \caption{Hadoop 2.0 stack with Spark (based on~\cite{big_data_working_group_big_2014}).}
    \label{fig:hadoop_stacks}
\end{figure}

Hadoop is the most widespread solution among the general-purpose big data platforms. Given its importance, we provide some details about its components in Fig.~\ref{fig:hadoop_stacks}, considering Hadoop v2.0 and Spark~\cite{big_data_working_group_big_2014}.

Hadoop v2.0 consists of the Hadoop kernel, MapReduce and the Hadoop Distributed File System (HDFS). YARN is the default resource manager, providing access to cluster resources to several competing jobs. Other resource managers (e.g., Mesos) and execution engines (e.g., TEZ) can be used too, e.g., for providing resources to Spark.

Spark has been introduced in Hadoop v2.0 onward aiming to solve limitations in the MapReduce paradigm. Spark is based on data representations that can be transformed into multiple steps while efficiently residing in memory. In contrast, the MapReduce paradigm relies on basic operations (i.e., map/reduce) that are applied to data batches read and stored to disk. Spark has gained momentum in non-batch scenarios, e.g, iterative and real-time big data applications, as well as in batch applications that cannot be solved in a few stages of map/reduce operations.

Several high-level language and systems, such as Google's Sawzall~\cite{pike_interpreting_2005}, Yahoo's Pig Latin~\cite{gates_building_2009}, Facebook's Hive~\cite{thusoo_hive_2009}, and Microsoft's SCOPE~\cite{chaiken_scope_2008} have been proposed to run on top of Hadoop. Moreover, several libraries such as Mahout~\cite{owen_mahout_2011} over MapReduce and MLlib over Spark have been introduced by the community to solve problems or fill gaps in the original Hadoop ecosystem. Finally, the ecosystem has been complemented with tools targeting specific processing models, such as GraphX and Spark Streaming, which support graph and stream processing, respectively.

%\footnote{Note that batch processing is a special case of streaming processing with larger update interval. In the streaming model, data instances arrive, and algorithms must process them under stringent time and space constraints. Hence, streaming algorithms often rely on approximations. These requirements challenge the traditional MapReduce and Spark models, thus calling for solutions such as Spark Streaming.}

Besides the Apache Hadoop distributions, proprietary platforms offer different features for data processing and cluster management. Some of such solutions include Cloudera CDH,\footnote{\url{http://www.cloudera.com/downloads/cdh/5-8-2.html}} Hortonworks HDP,\footnote{\url{http://hortonworks.com/products/data-center/hdp/}} and MapR Converged Data Platform.\footnote{\url{https://www.mapr.com/products/mapr-converged-data-platform}}

\section{Categorizing \ac{NTMA} applications}\label{sec:intro_tma}

\subsection{Taxonomy}

We rely on a subset of the taxonomy of network management applications found in~\cite{boutaba_comprehensive_2018} to categorize papers handling \ac{NTMA} with big data approaches. The previous work lists eight categories, which are defined according to the final purpose of the management application, namely: (i)~traffic prediction, (ii)~traffic classification, (iii)~fault management, (iv)~network security, (v)~congestion control, (vi)~traffic routing, (vii)~resource management, and (viii)~QoS/QoE management.

We only survey works that fit on the first four categories for two reasons. First, whereas the taxonomy in~\cite{boutaba_comprehensive_2018} is appropriate for describing network management applications, the level of dependence of such applications on \ac{NTMA} varies considerably. Traffic routing and resource management seem less dependent on large-scale measurements than traffic prediction and security, for example. Second, the literature on the use of big data approaches for congestion control, traffic routing, resource management, and QoS/QoE management is almost nonexistent by the time of surveying. We conjecture that either large-scale datasets targeting problems in these categories are not available, or researchers have not identified potential on applying big data approaches in those scenarios. We thus ignore the works using big data approaches for those cases.

\subsection{\ac{NTMA} applications}

Tab.~\ref{tbl:ntma_cats} shows the categories used in our survey. Next, we list examples of \ac{NTMA} applications in these categories.

\begin{table}[t!]
\caption{Categories of \ac{NTMA} applications~\cite{boutaba_comprehensive_2018}.}
\label{tbl:ntma_cats}
    \resizebox{\columnwidth}{!}{
    \begin{tabular}{l|l}
        Category               & Description \\
        \hline
        Traffic prediction     & \begin{tabular}[c]{@{}l@{}}Maintenance and planning of networks.\\ Historically achieved through time series forecasting.\end{tabular} \\\hline
        Traffic classification & \begin{tabular}[c]{@{}l@{}}Categorize and recognize traffic flows for different \\ objectives, e.g., QoE, security etc.\end{tabular} \\\hline
        Fault management       & \begin{tabular}[c]{@{}l@{}}Predict and isolate faults \\and unwanted behaviors in networks.\end{tabular}\\\hline
        Network security       & \begin{tabular}[c]{@{}l@{}}Protect the network, react to (and prevent from)\\ malicious activities and general attacks.\end{tabular}\\\hline
    \end{tabular}
}
\end{table}

\subsubsection{Traffic prediction}

Traffic prediction consists of estimating the future status of network links. It serves as a building block for traffic engineering, helping to define, for example, the best moment to deploy more capacity in the network so to keep QoS levels.

Traffic prediction is often faced as a time-series forecasting problem~\cite{boutaba_comprehensive_2018}.
Here both classic forecasting methods (e.g., ARIMA or SARIMA methods) and machine learning (e.g., deep neural networks) are employed. The problem is usually formulated as the estimation of traffic volumes based on previous measurements in the same links. Changes in network behavior, however, make such estimations very complicated. New services or sudden changes in service configurations (e.g., the deployment of novel bandwidth-hungry applications) poses major challenges to traffic prediction approaches.

\subsubsection{Traffic classification}

Traffic classification aims at identifying services producing traffic~\cite{valenti_reviewing_2013}. It is a crucial step for managing and monitoring the network. Operators need information about services, e.g., to understand their requirements and their impact on the overall network performance.

Traffic classification used to work well by simply inspecting information in network and transport protocols. For instance, Internet services used to be identified by simply inspecting TCP/UDP port numbers. However, traffic classification is no longer a simple task~\cite{nguyen_survey_2008}. First, the number of Internet services is large and continues to increase. Second, services must be identified by observing little information seen in the network. Third, little information remains visible in packets, since a major share of the Internet services run on top of a handful of encryption protocols (e.g., HTTPS over TCP). At last, Internet services are dynamic and constantly updated.

Many approaches have been proposed to perform traffic classification. They can be grouped according to the strategy used to classify the packets:
(i) Packet inspection analyzes the content of packets searching for pre-defined messages~\cite{bujlow_independent_2015} or protocol \emph{fingerprints};
(ii) Supervised machine learning methods extract features from the traffic and, in a training phase, build models to associate feature values to the services;
(iii) Unsupervised machine learning methods cluster traffic without previous knowledge on the services. As such, they are appropriate to explore services for which training data is unavailable~\cite{erman_traffic_2006}.
(iv) Behavioral methods identify services based on the behavior of end-nodes~\cite{karagiannis_blinc_2005}. The algorithms observe traffic to build models for nodes running particular services. The models describe, for instance, which servers are reached by the clients, with which data rate, the order in which servers are contacted, etc.

\subsubsection{Fault management}

Fault management is the set of tasks to predict, detect, isolate, and correct faults in networks. The goal is to minimize downtime. Fault management can be proactive, e.g., when analytics predict faults based on measurements to avoid disruptions, or reactive, e.g., when traffic and system logs are evaluated to understand ongoing problems. In either case, a key step in fault management is the localization of the root-cause of problems~\cite{boutaba_comprehensive_2018}.

In large networks, diverse elements may be impacted by faults: e.g., a failed router may overload other routes, thus producing a chain of faults in the network. Diverse network elements will produce system logs related to the problem, and the behavior of the network may be changed in diverse aspects. Detecting malfunctioning is often achieved by means of anomaly detection methods that identify abnormal behavior in traffic or unusual events in system logs. Anomalies, however, can be caused also by security incidents (described next) or normal changes in usage patterns. Analytics algorithms often operate evaluating traffic, system logs, and active measurements in conjunction, so to increase the visibility of the network and easy the identification of root-causes of problems.

\subsubsection{Security}

Many \ac{NTMA} applications have been proposed for assisting cyber-security~\cite{liao_intrusion_2013}. The most common objective is to detect security flaws, virus, and malware, so to isolate infected machines and take countermeasures to minimize damages. Roughly speaking, there are two main approaches when searching for malicious network activity: (i)~based on attack signatures; (ii)~based on anomaly detection.

Signature-based methods build upon the idea that it is possible to define fingerprints for attacks. A monitoring solution inspects the source traffic/logs/events searching for (i)~known messages exchanged by viruses, malware or other threats; or (ii) the typical communication patterns of the attacks -- i.e., similar to behavioral traffic classification methods. Signature-based methods are efficient to block well-known attacks that are immutable or that mutate slowly. These methods however require attacks to be well-documented.

Methods based on anomaly detection~\cite{bhuyan_network_2014,garcia-teodoro_anomaly-based_2009} build upon the assumption that attacks will change the behavior of the network. They build models to summarize the \emph{normal} network behavior from measurements. Live traffic is then monitored and alerts are triggered when the behavior of the network differs from the baseline. Anomaly detection methods are attractive since they allow the early detection of unknown threats (e.g., zero-day exploits). These methods, however, may not detect stealth attacks (i.e., false negatives), which are not sufficiently large to disturb the network. They sometimes suffer from large numbers of false positives too.

\subsection{Big data challenges in \ac{NTMA} applications}

We argue that \ac{NTMA} applications belonging to the categories above can profit from big data approaches. Processing such measurements poses the typical big data challenges (i.e., the ``5-Vs''). We provide some examples to support the claim.

Considering \emph{volume} and \emph{velocity} and taking traffic classification as an example: it has to be performed on-the-fly (e.g., on packets), and the input data are usually large. We will see later that researchers rely on different strategies to perform traffic classification on high-speed links, with some works applying algorithms to hundreds of Gbps streams.

Consider then \emph{variety}. As more and more traffic goes encrypted, algorithms to perform traffic classification or fault management, for example, have poorer information to operate. Modern algorithms rely on diverse sources -- e.g., see~\cite{trevisan_towards_2016} that combines DNS and flow measurements. Anomaly detection, as another example, is usually applied to a variety of sources too (e.g., traffic traces, routing information, etc), so to obtain diverse signals of anomalies.

In terms of \emph{veracity}, we again cite cyber-security. Samples of attacks are needed to train classifiers to identify the attacks on real networks. Producing such samples is a challenging task. While simple attacks can be reproduced in laboratory, elaborate attacks are hard to reproduce -- or, worst, are simulated in unrealistically ways -- thus limiting the analysis.

Finally, \emph{value} is clearly critical in all the above applications -- e.g., in cyber-security, a single attack that goes undetected can be unbearable to the operation of the network.

\section{Data management for NTMA}
\label{sec:big_data_for_NTMA}

We now survey how the data management steps (cf. Fig.~\ref{fig:KDD_process}) are performed in NTMA applications. Big data analytics for NTMA are instead covered in Sect.~\ref{sec:big_data_analytics_for_NTMA}.

\subsection{Data collection}
\label{sub_sec:capturing_NTMA_Big-Data}

Measuring the Internet is a big data challenge. There are more than half a million networks, 1.6 billion websites, and more than 3 billion users, accessing more than 50 billion web pages, i.e., exchanging some zettabytes per year. At no surprise, the community has spent much work in designing and engineering tools for data acquisition at high-speeds; some of them are discussed in~\cite{sakr_big_2018}. Here the main challenges addressed by the community seem to be the scalability of data collection systems and how to reduce data volumes at the collection points without impacting the data quality.

Measuring the Internet can be accomplished coarsely in two means: (i) active measurements, and (ii) passive measurements. The former indicates the process of injecting data into the network and observing the responses. It is a not scalable process, typically used for troubleshooting. Passive measurements, on the contrary, build on the continuous observation of traffic, and the extraction of indexes in real-time.

A network \emph{exporter} captures the traffic crossing monitoring points, e.g., routers aggregating several customers. The exporter sends out a copy of the observed network packets and/or traffic statistics to a \emph{collector}. Data is then saved in \emph{storage} formats suitable for NTMA applications. \emph{Analysis} applications then access the data to extract knowledge.

The components of this architecture can be integrated into a single device (e.g., a router or a dedicated network monitor) or deployed in a distributed architecture. We will argue later that big data already emerges since the first stage of the NTMA process and, as such, distributed architectures are common in practical setups. The large data rate in the monitoring points has pushed researchers and practitioners into the integration of many pre-processing functionalities directly into the exporters. The most prominent example is perhaps flow-based monitoring, in which only a summary of each traffic flow is exported to collectors.
Next, we provide a summary of packet- and flow-based methods used to collect data for NTMA applications.

\subsubsection{Packet-based methods}

\begin{table*}[!htb]
\small
\centering
\caption{Size of real packet traces captured at different vantage points with different methodologies.}
\begin{tabular}{|x{8cm}|x{2cm}|x{2cm}|x{3cm}|x{0cm}}
  \cline{1-4}
    Description & Duration & \texttt{pcap} (GB) & Headers up to L4 (\%) & \\[0.5ex]
  \cline{1-4}
    Full packets of 20~k ADSL users (morning) & 60~mins & 675 & 6.8 & \\ [.5ex]
    Full packets of 15~k Campus users (morning) &  100~mins & 913 & 6.3 & \\ [.5ex]
    Only 5~kB or 5 packets per flow, 20~k ADSL users & 1~day & 677 & 22.5 & \\ [.5ex]
  \cline{1-4}
\end{tabular}
\label{tab:packet_volumes}
\end{table*}

Analyzing packets provides the highest flexibility for the NTMA applications, but also requires a significant amount of resources. Deep Packet Inspection (DPI) means to look into, and sometimes export, full packet contents that include application headers and payload. The data rate poses challenges (i.e., velocity), which can be faced using off-the-shelf hardware~\cite{trevisan_traffic_2017}, provided hardware support is present~\cite{intel_data_2014}. Indeed, there exist technologies to perform DPI at multiple Gbit/s, while also saving the packets to disk~\cite{deri_10_2013}.

The network monitoring community has proposed multiple alternatives to avoid the bottlenecks of packet-based methods. The classical approach is to perform pre-processing near the data collection. Filtering and sampling are usually set on collection points to export only a (small) subset of packets that pass the filtering and sampling rules. Other ad-hoc transformations may be employed too, e.g., the exporting of packet headers only, instead of full packet contents. In a similar direction, authors of~\cite{maier_enriching_2008} propose to collect only the initial packets of each flow, since these packets are usually sufficient for applications such as traffic classification.

As an illustration, Tab.~\ref{tab:packet_volumes} describes the volume of packet traces captured in some real deployments. The first two lines report the size of packet traces capture at (i)~an ISP network where around 20\,k ADSL customers are connected; (ii)~a campus network connecting around 15\,k users. Both captures have been performed in morning hours and last for at least 1 hour.  More than half of TB is saved in both cases. The table also shows that the strategy of saving only packet headers up to the transport layer only partially helps to reduce the data volume -- e.g., around 45\,GB of headers per hour would still be saved for the ISP trace. Finally, the last line shows the size of a full day of capture in the ISP network with a setup similar to~\cite{maier_enriching_2008} (i.e., saving only 5 packets or 5\,kB per flow). Whereas the data volume is reduced significantly, more than 600\,GB of \texttt{pcaps} are saved per day.

Nowadays, DPI is challenged by encryption as well as by restrictive privacy policies~\cite{fuchs_implications_2012}. Alternatives to performing DPI on encrypted traffic have been presented~\cite{sherry_blindbox_2015}.

\subsubsection{Flow-based methods}

Flow-based methods process packets near the data collection, exporting only summaries of the traffic \emph{per flow}~\cite{hofstede_flow_2014}. A network flow is defined as a sequence of packets that share common characteristics, identified by a \emph{key}. NTMA applications that analyze flow records have lower transmission requirements since data is aggregated. Data privacy is better protected and issues related to encryption are partially avoided. Nevertheless, there is an unavoidable loss of information, which makes the analysis more limited. In 2013, Cisco estimated that approximately 90\% of network traffic analyses are flow-based, leaving the remaining 10\% for packet-based approaches~\cite{patterson_netflow_2013}.

Diverse flow monitoring technologies have been proposed. Cisco NetFlow, sFlow and IPFIX are the most common ones. NetFlow has been widely used for a variety of NTMA applications, for example for network security~\cite{nickless_combining_2000}. sFlow relies heavily on packet sampling (e.g., sampling 1 every 1\,000 packets). This may hurt the performance of many NTMA applications, even if sFlow is still suitable for detecting attacks for example. Finally, IPFIX (IP Flow Information Export protocol) is the IETF standard for exporting flow information~\cite{hofstede_flow_2014}. IPFIX is flexible and customizable, allowing one to select the fields to be exported. IPFIX is used for a number of NTMA applications, such as the detection of SSH attacks~\cite{hofstede_ssh_2014}.

Even flow-based monitoring may produce very large datasets~\cite{sakr_big_2018}. For illustration, Tab.~\ref{tbl:data_compression} lists the volume of flow-level information exported when monitoring 50~Gbit/s on average. The table is built by extrapolating volumes reported by~\cite{hofstede_flow_2014} for flow exporters in a real deployment. The table includes Cisco's NetFlow~v5 and NetFlow~v9 as well as IPFIX.

\begin{table}[!t]
  \renewcommand{\arraystretch}{1.3}
  \caption{Estimated measurement streams for a flow exporter observing 50~Gbit/s on average (source~\cite{hofstede_flow_2014,sakr_big_2018}).}
  \label{tbl:data_compression}
  \centering
    \begin{tabular}{| c | c | c | c |} \hline
      \textbf{Sampling rate} & \textbf{Protocol}           & Packets/s  & Bits/s \\ \hline
      1:1                    & NetFlow~v5                  &  4.1 k     & 52 M  \\ \hline
      1:1                    & \multirow{2}{*}{NetFlow~v9}  & 10.2 k     & 62 M  \\ \cline{1-1} \cline{3-4}
      1:10                   &                             &  4.7 k     & 27 M  \\ \hline
      1:1                    & IPFIX                       & 12.5 k     & 75 M  \\ \hline
    \end{tabular}
\end{table}

Flow-based NTMA applications would still face more than 70~Mbit/s if IPFIX is chosen to export data. Since these numbers refer to a single vantage point, NTMA applications that aggregate multiple vantage points may need to face several Gb/s of input data. Finally, the table reports data speeds when employing sampling (see 1:10 rate). Sampling reduces the data volume considerably but limiting NTMA applications.

At last, NTMA applications often need to handle historical data. The storage needed to archive the compressed flow data from the vantage point used as an example in Tab.~\ref{tab:packet_volumes} grows linearly over time, and the archival consumes almost 30~TB after four years~\cite{sakr_big_2018} of archival.

\subsection{Data ingestion}\label{sub_sub_sec:big_data_acquisition}

The previous step is the KDD step that depends the most on the problem domain since the way data is acquired varies a lot according to the given scenario. Not a surprise, generic big data frameworks mostly provide tools and methods for \emph{transporting} raw data from its original sources into pre-processing and analytics pipelines. The process of transporting the data into the frameworks is usually called \emph{data ingestion}. Since data at its source is still unprocessed, handling such raw data may be a huge challenge. Indeed, large data streams or a high number of distributed collectors will produce a deluge of data to be parsed and pre-processed in subsequent stages.

Several generic tools for handling big data ingestion can be cited:
(i)~\emph{Flume},\footnote{\url{https://flume.apache.org}} a distributed system to collect logs from different sources;
(ii)~\emph{Sqoop},\footnote{\url{http://sqoop.apache.org}} which allows to transmit data between Hadoop and relational databases;
(iii)~\emph{Kafka},\footnote{\url{https://kafka.apache.org}} a platform that provides a distributed publish-subscribe messaging system for online data streaming; among others.
These tools focus on scalability -- e.g., they allow multiple streams to be processed in a distributed fashion. Once data is delivered to the framework, pre-processing can take place on the streams.

Considering NTMA, few solutions have been proposed to ingest traffic measurements into big data frameworks. Here the main challenges seem to be the transport of data in different formats and from various sources into the frameworks. The research community has made interesting progresses in this front, and we cite two examples.

Apache Spot\footnote{\url{http://spot.incubator.apache.org}} is an open platform that integrates modules for different applications in the network security area. It relies on Kafka for performing data ingestion. Users and developers have to provide Spot with python scripts that parse the original measurements into a pre-defined, but flexible, format. Both the original data and converted records are loaded and stored in the big data framework.

A promising project is PNDA (Platform for Network Data Analysis).\footnote{\url{http://pnda.io/overview}} Developed by the Linux Foundation, it is an open-source platform capable of managing and grouping together different data sources. It then performs analysis on such data sources. It does not force data in any specific schema and it allows the integration of other sources, producing custom code for the analysis stage. It makes use of standard tools in big data frameworks, including Kafka for data ingestion. Here, a number of plugins for reading data in virtually all popular NTMA formats is available on the project website.

Spot and PNDA experiences clearly show that ingesting NTMA data into generic frameworks requires some effort to plugin the NTMA formats into the frameworks. Once such plugins exist, standard ingestion tools like Kafka can be employed. The availability of open source code (e.g., PNDA plugins) makes it easier to integrate these functionalities in other NTMA applications.

\subsection{Data storage}
\label{sub_sec:ntma_storage}

In theory, generic store systems of big data frameworks can be used with NTMA too. In practice, a number of characteristics of the big data frameworks complicate the storage of NTMA data. In order to ease the understanding of these challenges, we first provide some background in generic big data storage systems. Then, we evaluate how the NTMA community is employing these systems in NTMA applications.

\subsubsection{Generic systems}

\begin{figure}[t!]
    \centering
    \includegraphics[width=1.025\linewidth]{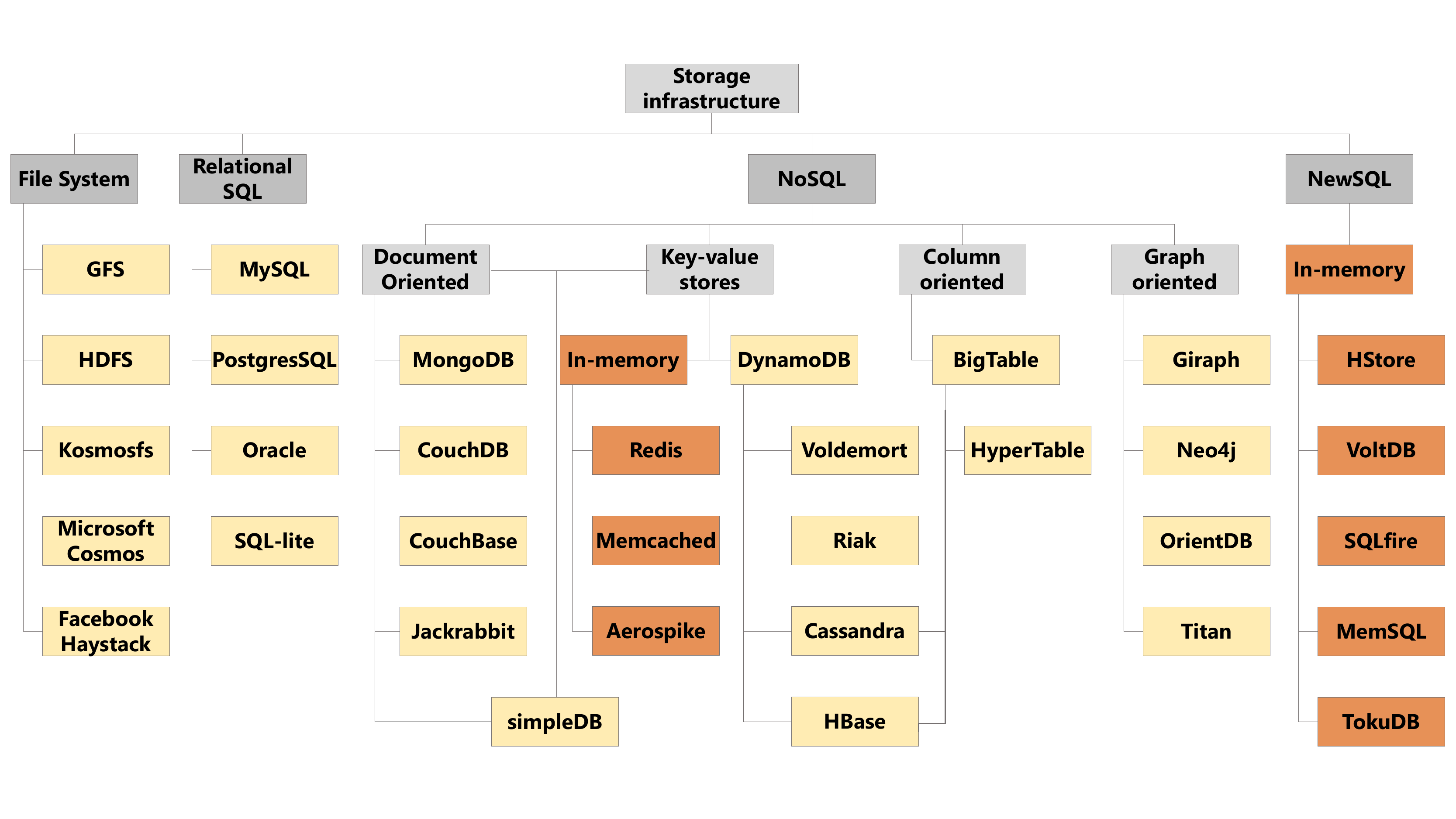}
    \caption{Generic big storage systems (based on~\cite{chen_big_2014,big_data_working_group_big_2014}).}
    \label{fig:storage_management_systems}
\end{figure}

Fig.~\ref{fig:storage_management_systems} reproduces a taxonomy of big data storage and management systems~\cite{chen_big_2014,big_data_working_group_big_2014}. Colors represent the media (i.e., memory or persistent media) primarily exploited by the system.

Most big data storage systems focus on horizontal scalability, i.e., growing the capacity of the system across multiple servers, rather than upgrading a single server to handle increasing data volumes. This approach results in large distributed systems, which carry many risks, such as node crashes and network failures. Consistency, availability, and fault tolerance are therefore of large concern in big data storage~\cite{big_data_working_group_big_2014, hu_toward_2014}.

In terms of file systems, Google has pioneered the development by the implementation of the Google File System (GFS)~\cite{ghemawat_google_2003}. GFS was built with horizontal scalability in mind, thus running on commodity servers and providing fault tolerance and high performance. Colossus~\cite{mckusick_gfs_2009}, the successor of GFS and Facebook Haystack~\cite{beaver_finding_2010} are other examples. Open source derivatives of GFS appeared later, including Apache HDFS and Kosmosfs.

On a higher abstraction level, relational databases suffer performance penalties with large datasets. NoSQL databases~\cite{han_survey_2011} are alternatives that emerged for such scenarios. NoSQL databases scale better than the relational ones by overcoming the intrinsic rigidity of database schemas. NoSQL databases adopt different physical data layouts:
(i)~\emph{Key-value} databases store data in sets of key-value pairs organized into rows. Examples include Amazon's Dynamo~\cite{decandia_dynamo_2007} and Memcached;\footnote{\url{http://memcached.org}}
(ii)~\emph{column-oriented} databases (inspired on Google's BigTable~\cite{chang_bigtable_2008}) store data by column instead of by row. Examples are Cassandra~\cite{lakshman_cassandra_2010} and HBase;\footnote{\url{https://hbase.apache.org/}}
(iii)~\emph{document oriented} databases store data in documents uniquely identified by a key. An example is MongoDB;\footnote{\url{https://www.mongodb.com/}}
(iv)~\emph{Graph oriented} databases store graphs -- i.e., nodes and edges. They impose a graph schema to the database, but profiting from it to implement efficient operations in graphs. An example is Titan.\footnote{\url{http://titan.thinkaurelius.com/}}

Finally, NewSQL systems aim at providing a similar performance of NoSQL databases, while maintaining properties of relational systems~\cite{stonebraker_newsql_2011}, e.g., relational data models and SQL queries~\cite{cattell_scalable_2011}. An example is Google Spanner.\footnote{\url{https://cloud.google.com/spanner/}}

\subsubsection{NTMA storage in big data frameworks}

Being the Internet a distributed system, a key problem is how to archive measurements in a centralized data store. Here no standard solution exists, despite multiple attempts to provide scalable and flexible approaches~\cite{sacerdoti_wide_2003,trammell_mplane_2014}. The measurements are usually collected using ad-hoc platforms and exported in formats that are not directly readable by big data frameworks. Therefore, both special storage solutions and/or additional data transformation steps are needed.

For example, \texttt{libpcap}, \texttt{libtrace}, \texttt{libcoral}, \texttt{libnetdude} and \texttt{Scapy} are some libraries used for capturing packets. These libraries read and save traces using the \texttt{pcap} format (or \texttt{pcap-ng}). Distributed big data file and database systems, such as HDFS or Cassandra, are generally unable to read \texttt{pcap} traces directly, or in parallel, since the \texttt{pcap} traces are not separable -- i.e., they cannot be easily split into parts. One would still need to reprocess the traces sequentially when reading data from the distributed file systems. A similar problem emerges for formats used to store flow-based measurements. For example, \texttt{nfdump} is a popular format used to store NetFlow measurements. While files are split into parts by the collector according to pre-set parameters (e.g., saving one file every five minutes), every single file is a large binary object. Hadoop-based systems cannot split such files into parts automatically.

A number of previous works propose new approaches to overcome these limitations: (i)~loading the measurements in original format to the distributed system, while extending the frameworks to handle the classic formats; (ii)~proposing new storage formats and tools for handling the big network data; (iii)~transforming the network data and importing it into conventional big data storage systems (discussed in Sect.~\ref{sub_sec:ntma_pre-processing}).

Lee and Lee~\cite{lee_toward_2013} have developed a Hadoop API called \emph{PcapInputFormat} that can manage IP packets and NetFlow records in their native formats. The API allows NTMA applications based on Hadoop to seamlessly process \texttt{pcap} or NetFlow traces. It thus allows traces to be processed without any previous transformation while avoiding performance penalties of reading files sequentially in the framework. A similar direction is followed by Ibrahim et al. \cite{ibrahim_study_2015}, who develop traffic analysis algorithms based on MapReduce and a new input API to explore \texttt{pcap} files on Hadoop.

In \cite{nagele_large-scale_2011}, Nagele faces the analysis of \texttt{pcap} files in a fast and scalable way by implementing a java-based hadoop-pcap library. The project includes a Serializer/Deserializer that allows Hive to query \texttt{pcaps} directly. Authors of~\cite{tazaki_matatabi_2014} use the same Serializer/Deserializer to build a Hadoop-based platform for network security that relies on sFlow, Netflow, DNS measurements and SPAM email captures.

Noting a similar gap for processing BGP measurements, authors of \cite{orsini_bgpstream_2016} introduce BGPStream. BGPStream provides tools and libraries that connect live BGP data sources to APIs and applications. BGPStream can be used, for instance, for efficiently ingesting on-line data into stream processing solutions, such as Spark Streaming.

All these APIs are however specific, e.g., to HDFS, \texttt{pcap} or NetFlow. Thus, similar work has to be performed for each considered measurement format or analytics framework.

Some authors have taken a distinct approach, proposing new storage formats and solutions more friendly to the big network measurements. Authors of \cite{bar_large-scale_2014} propose DBStream to calculate continuous and rolling analytics from data streams. Other authors have proposed to extend \texttt{pcap} and flow solutions both to achieve higher storage efficiency (e.g., compressing traces) and to include mechanisms for indexing and retrieving packets efficiently~\cite{fusco_net-fli_2010,fusco_pcapindex_2012}. These solutions are all built upon key ideas of big data frameworks, as well as key-value or column-oriented databases. They are however specialized to solve network monitoring problems. Yet, the systems are generally centralized, thus lacking horizontal scalability.

\subsection{Pre-processing}
\label{sub_sec:ntma_pre-processing}

NTMA algorithms usually operate with \emph{feature vectors} that describe the instances under study -- e.g., network flows, contacted services, etc. We, therefore, consider as pre-processing all steps to convert raw data into feature vectors.

Some papers overcome the lack of storage formats by transforming the data when ingesting it into big data frameworks. Similarly, a set of features must be extracted for performing NTMA, both for packet and flow-based analysis. Next, we review works performing such pre-processing tasks for NTMA.

\subsubsection{Transformations}

A popular approach to handle the big network measurements is to perform further transformations after the data leaves the collection point. Either raw \texttt{pcap} or packets pre-processed at the collection point (e.g., sampled or filtered) are passed through a series of transformations before being loaded into big data frameworks.

Authors of \cite{wullink_entrada_2016} use an extra pre-processing step to convert measurements from the original format (i.e., \texttt{pcap} in HDFS) into a query-optimized format (i.e., Parquet). This conversion is shown to bring massive improvements in performance, but only pre-selected fields are loaded into the query-optimized format. Authors of \cite{samak_scalable_2012} propose a Hadoop-based framework for network analysis that first imports data originally in \mbox{perfSONAR} format into Apache Avro. Authors of \cite{lee_internet_2010} process flow logs in MapReduce, but transforming the original binary files into text files before loading them into HDFS.

Marchal et al. \cite{marchal_big_2014} propose ``a big data architecture for large-scale security monitoring'' that uses DNS measurements, NetFlow records and data collected at honeypots. Authors take a hybrid approach: they load some measurements into Cassandra while also deploying Hadoop APIs to read binary measurement formats directly (e.g., \texttt{pcaps}). The performance of the architecture is tested with diverse big data frameworks, such as Hadoop and Spark. Similarly, Spark Streaming is used to monitor large volumes of IPFIX data in \cite{cermak_real-time_2016}, with the IPFIX collector passing data directly to Spark in JSON format.

In \cite{li_supervised_2013} sFlow records of a large campus network are collected and analyzed on a Hadoop system in order to classify host behaviors based on machine learning algorithms. sFlow data are collected, fields of interest are extracted and then ingested into Cassandra using Apache Flume. Sarlis et al. propose a system for network analytics based on sFlow and NetFlow (over Hadoop or Spark) that achieves 70\% speedup compared to basic analytics implementations with Hive or Shark~\cite{sarlis_datix_2015}. Measurements are first transformed into optimized partitions that are loaded into HDFS and HBASE together with indexes that help to speed up the queries.

An IPFIX-based lightweight methodology for traffic classification is developed in \cite{murgia_lightweight_2016}. It uses unsupervised learning for word embedding on Apache Spark, receiving as input ``decorated flow summaries'', which are textual flow summaries augmented with information from DNS and DHCP logs. Finally, specifically for big data scenarios, Cisco has published a comprehensive guide for network security using Netflow and IPFIX \cite{santos_network_2015}. Cisco presents OpenSOC, an integral solution to protect again intrusion, zero-day attacks and known threats in big data frameworks. OpenSOC includes many functionalities to parse measurements and load them into big data frameworks using, for example, Apache Flume and Kafka.

All these works reinforce the challenge posed by the lack of NTMA formats friendly to big data frameworks. At the one hand, transformations boost analysis performance, and performance seems to be the main focus of the community so far. At the other hand, information may be lost in transformations. Eventually, data replicated in many formats increase redundancy and make harder integration with other systems.

\subsubsection{Feature engineering}

Several libraries exist to perform feature extraction and selection in big data frameworks. While many of them are domain-specific, generic examples are found in Spark ML and Spark MLlib.\footnote{https://spark.apache.org/mllib/} Instead, the research about feature extraction and selection in NTMA is scarce in general. Here the main challenges seem to be the lack of standard or large-accepted features for each type of NTMA applications. Minimally, the community lacks ways to compare and benchmark systems and features in a systematic way.

Most works either refer to old datasets, e.g.,~\cite{chen_survey_2006,li_building_2009,amiri_mutual_2011}. The work in~\cite{iglesias_analysis_2015} studies traffic features in classic datasets for attack/virus detection and DM/ML testing (e.g., DARPA~\cite{mit_lincoln_laboratories_darpa_1999} dataset). Such datasets have been criticized and their use discouraged~\cite{mchugh_testing_2000}. Authors consider whether the features are suitable or not for anomaly detection, showing that features present high correlation, thus mostly being unnecessary.

Abt et al.~\cite{abt_performance_2013} study the selection of NetFlow features to detect botnet C\&C communication, achieving accuracy and recall rates above 92\%. In~\cite{kim_combined_2004} Netflow features undergo feature selection for the case of DDoS detection. In~\cite{valenti_identifying_2011} IPFIX records are used in feature selection processes, obtaining a set of key features for the classification of P2P traffic. All these papers handle relatively small datasets. Few authors rely on large datasets~\cite{arzani_taking_2016}, but instead propose features that are finely tailored to the specific problem at hand. In a nutshell, each work proposes a custom feature engineering, with no holistic solution yet.

\section{Big data analytics for NTMA}\label{sec:big_data_analytics_for_NTMA}

The next step of the NTMA path is data analysis. We recall that we do not intend to provide an exhaustive survey on general data analytics for NTMA. In particular, here we focus on how the main analytics methods are applied to \emph{big} NTMA datasets. Detailed surveys of other NTMA scenarios are available, e.g., in~\cite{bhuyan_network_2014,garcia-teodoro_anomaly-based_2009,liao_intrusion_2013,nguyen_survey_2008}.

As for data management, generic frameworks exist and could be used for NTMA. We next briefly summarize them. After it, we dig into the NTMA literature.

\subsection{Generic big data frameworks}\label{sec:data_analysis}

\begin{figure}[t!]
	\centering
	\includegraphics[width=1.15\linewidth]{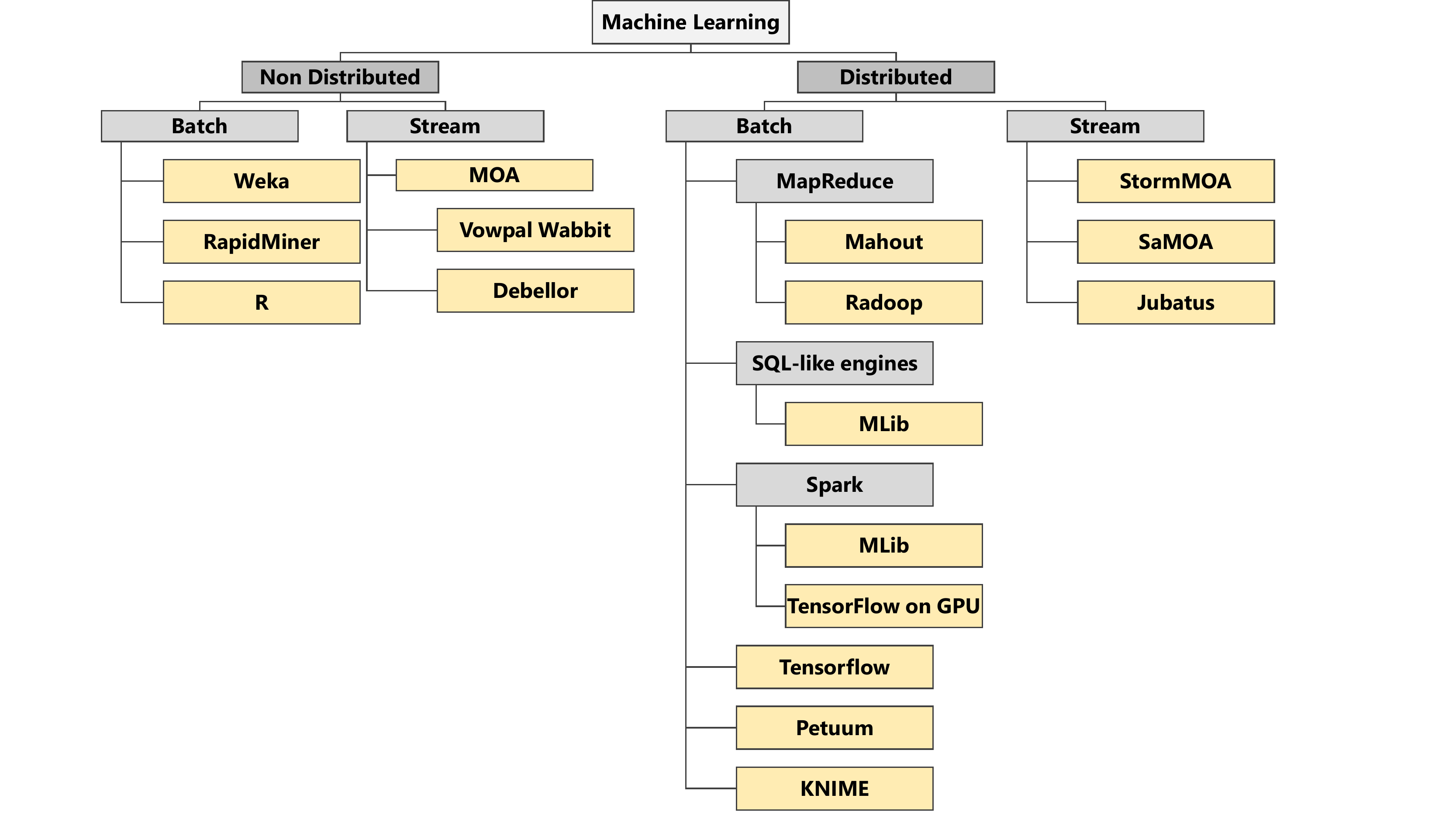}
	\caption{Machine learning for big data (based on~\cite{bifet_big_2014}).}
	\label{fig:taxonomy-of-machine-learning-frameworks}
\end{figure}

Several taxonomies have been proposed to describe analysis algorithms. Algorithms are roughly classified as (i) statistical or (ii) machine learning. Machine learning are further categorized as supervised, unsupervised and semi-supervised, depending on the availability of ground truth and how data is used for training. Novel approaches are also often cited, such as deep neural networks and reinforcement learning.

Many challenges emerge when applying such algorithms to big data, due to the large volumes, high dimension etc. Some machine learning algorithms simply do not scale linearly with the input size, requiring lots of resources for processing big data sets. These problems are usually tackle by (i) pre-processing further the input to reduce its complexity; (ii) parallelizing algorithms, sometimes replacing exact solutions by more efficient approximate alternatives.

Several approaches have been proposed to parallelize statistical algorithms~\cite{hastie_elements_2009} or to scale machine learning algorithms~\cite{raykar_fast_2007,sun_sparse_2010,jiang_cross-domain_2008}. Parallel version of some statistical algorithms are presented in~\cite{bennett_numerically_2009}. P\'{e}bay et al.~\cite{pebay_design_2011} provide a survey of parallel statistics. Parallel neural networks are described in~\cite{mikolov_strategies_2011,byungik_ahn_neuron_2012,yuan_privacy_2014}. Parallel training for deep learning are covered in~\cite{bengio_learning_2009,le_building_2013,ciresan_multi-column_2012}.

Frameworks do exist to perform such analyses on big data. Fig.~\ref{fig:taxonomy-of-machine-learning-frameworks} reproduces a taxonomy of machine learning tools able to handle large data sets~\cite{bifet_big_2014}. Both non-distributed (e.g., Weka and R) and distributed (e.g., Tensorflow and MLlib) alternatives are popular. Additional challenges occur with streaming data, since algorithms must cope with strict time constraints. We see in the figure that tools targeting these scenarios are also available (e.g., StormMOA).

In terms of NTMA analytics, generic framework implementing algorithms that can scale to big data could be employed too, naturally. Next we explore the literature to understand whether big data approaches and frameworks are actually employed in NTMA.

\subsection{Literature categorization}

In our examination, we focus on understanding the depth of the application of big data techniques in NTMA. In particular Tab.~\ref{tab:bd_NTMA_summary} evaluates each work under the following perspectives:

\begin{itemize}

\item We check whether works face large data \textit{volumes}. We arbitrarily define thresholds to consider a dataset to be big data: All works handling data larger than tens of GBs or works handling backbone network use cases. Similarly, we consider big data volumes when the study covers periods of months or years if dataset size is not specified.

\item We verify if popular \textit{frameworks} are used, i.e., \textit{Spark}, \textit{Hadoop}, \textit{MapReduce}; we accept also custom implementations that exploit the same paradigms.

\item We check if \textit{machine learning} is used for NTMA.

\item We verify if \textit{big data platforms} are used in the ML process (see Fig.~\ref{fig:taxonomy-of-machine-learning-frameworks}).

\item We check \textit{velocity}, i.e., if works leverage online analysis.

\item We address \textit{variety}, i.e., if authors use more than one data source in the analysis.

\end{itemize}

We leave out two of the 5 ``V's'', i.e., \textit{veracity} and \textit{value}, since it is cumbersome to evaluate them. For example, while some works discuss aspects of veracity (e.g., highlighting false positives of trained classifiers), rarely the veracity of the big data used as input in the analysis is evaluated.

Tab.~\ref{tab:bd_NTMA_summary} shows that big data techniques are starting to permeate NTMA applications. Network security attracts more big data research. In general, it is interesting to notice the adoption of machine learning techniques. However, observe the limited adoption of big data platforms for machine learning.

Next, we dig into salient conclusions of this survey.

\begin{table*}
   \centering
   \caption{Analyzed papers divided by application category (rows) and big data characteristics (columns).}
    \label{tab:bd_NTMA_summary}
   \begin{tabularx}{\textwidth}{|l|X|X|X|X|X|X|}
    \hline
    Category & Volume? & Big data framework? & ML based?	&  Big data ML? & Velocity - Online analysis? & Variety? \\
    \hline
     Traffic Prediction &
     % Volume
           \cite{fiadino_call_2017} \cite{shadi_hierarchical_2017} \cite{gonzalez_net2vec_2017}\cite{wassermann_netperftrace_2017} \cite{tian_tadoop_2015} &
    %

     %Framework
       \cite{tian_tadoop_2015}
     &
     %

     %AI
      \cite{fiadino_call_2017} \cite{shadi_hierarchical_2017} \cite{gonzalez_net2vec_2017}\cite{wassermann_netperftrace_2017}  \cite{tian_tadoop_2015} &
     %

    % AI B-D
     &

    %Velocity
     \cite{gonzalez_net2vec_2017} \cite{tian_tadoop_2015} &
     %

     % Variety
      \cite{fiadino_call_2017} \\
     \hline

     Traffic Classification &
  % Volume

    \cite{apiletti_selina_2016}\cite{garcia_efficient_2018}\cite{li_deep_2018}\cite{trevisan_awesome_2018}\cite{vassio_users_2017} &
    %Framework
     \cite{apiletti_selina_2016}\cite{li_deep_2018}\cite{casas_gml_2017} \cite{trevisan_awesome_2018}
     \cite{vassio_users_2017}&
     %AI
     \cite{apiletti_selina_2016} \cite{fiadino_grasping_2016}\cite{garcia_efficient_2018}\cite{zhao_data_2018}\cite{li_deep_2018} \cite{schiff_netslicer_2018} \cite{casas_gml_2017}\cite{trevisan_awesome_2018} &
     %AI Bigdata
     \cite{apiletti_selina_2016}\cite{trevisan_awesome_2018}
     &
     %Velocity
     \cite{apiletti_selina_2016}\cite{garcia_efficient_2018} \cite{fiadino_grasping_2016} &
     %Variety
     \\
     \hline

    % Traffic Routing &
    %  \cite{Comarela2013interdomain}  \cite{Geyer2018} \cite{polverini2018raw} &
    % \cite{Comarela2013interdomain} \cite{polverini2018raw}&
    % \cite{Geyer2018} &
    % &
    % &
    % \cite{Comarela2013interdomain}\\
    % %
    % \hline
    % Congestion Control & \cite{Zhang2016} & & \cite{Zhang2016}  & & & \\
    % %
    % \hline

    % Resource Management &
    % % Volume
    % \cite{trevisan_awesome_2018}  %\cite{vanhove2015live}&
    % % BD Framework
    % \cite{trevisan_awesome_2018}\cite{vanhove2015live} &
     % AI
    % \cite{trevisan_awesome_2018} \cite{kobayashi_mining_2018} %\cite{Yan2018} \cite{Blenk2017} \cite{otomo_finding_2018} &
     %AI with Big Data
    %  \cite{trevisan_awesome_2018} \cite{kobayashi_mining_2018} &
      %Velocity
    %  \cite{vanhove2015live}&
      %Variety
    %  \\
     %
    % \hline

    Fault Management &
    % Volume
     \cite{otomo_finding_2018} \cite{arzani_taking_2016} \cite{harper_cookbook_2018} \cite{vaarandi_unsupervised_2018} \cite{fontugne_hashdoop_2014} \cite{kasai_network_2016}\cite{putina_telemetry-based_2018} \cite{clemm_dna_2015}\cite{chandramouli_model-driven_2017}&
    % BD Framework
     \cite{fontugne_hashdoop_2014}\cite{clemm_dna_2015}\cite{clemm_dna_2015}
    &
    % AI/ML
    \cite{arzani_taking_2016} \cite{harper_cookbook_2018} \cite{vaarandi_unsupervised_2018} \cite{fontugne_hashdoop_2014}\cite{kasai_network_2016} \cite{putina_telemetry-based_2018}\cite{kobayashi_mining_2018}
    \cite{vassio_users_2017}&
    % AI/ML with big data
    \cite{fontugne_hashdoop_2014}\cite{kobayashi_mining_2018}&
    % Velocity
    \cite{fontugne_hashdoop_2014}\cite{kasai_network_2016}\cite{clemm_dna_2015}&
    % Variety
    \cite{kasai_network_2016}
    \cite{clemm_dna_2015}
    \\
    \hline

    % QoS/QoE Management &
    % \cite{Mestres2018} &
    % &
    % \cite{Mestres2018} &
    % &
    % &
    % \\
    % %
    % \hline

     Network Security &
     % Volume
      \cite{benzidane_toward_2016}
     \cite{rathore_hadoop_2016}  \cite{spina_snooping_2015}   \cite{huang_new_2016} \cite{mulinka_stream-based_2018}\cite{lee_detecting_2011} \cite{uwagbole_applied_2017} \cite{hameed_efficacy_2016}&
     % BD Framework
     \cite{benzidane_toward_2016}
     \cite{rathore_hadoop_2016}  \cite{shibahara_malicious_2017} \cite{spina_snooping_2015} \cite{li_world_2016} \cite{cogranne_detecting_2018} \cite{vanerio_ensemble-learning_2017} \cite{lee_detecting_2011} \cite{hameed_efficacy_2016}&
     % AI
     \cite{rathore_hadoop_2016}  \cite{shibahara_malicious_2017} \cite{spina_snooping_2015}  \cite{li_world_2016} \cite{cogranne_detecting_2018} \cite{frishman_cluster-based_2017} \cite{vanerio_ensemble-learning_2017} \cite{mulinka_stream-based_2018}
     \cite{uwagbole_applied_2017}&
     % AI w big data
    \cite{cogranne_detecting_2018}
    \cite{li_world_2016}\cite{tian_tadoop_2015}
     &
    % Velocity
     \cite{benzidane_toward_2016}
       \cite{mulinka_stream-based_2018} \cite{rathore_hadoop_2016} \cite{tian_tadoop_2015} \cite{hameed_efficacy_2016}&
      % Variety
        \\
     \hline
\end{tabularx}

\end{table*}

\subsection{Single source, early reduction, sequential analysis}
\label{sub_sec:analisys_seq}

From Tab.~\ref{tab:bd_NTMA_summary}, we can see how most of the works are dealing with the big volumes of data. This outcome is predictable since network traffic is one of the leading sources of big data. From the last column, variety, we can notice that the researchers infrequently use different sources together, limiting the analysis on specific use cases.

As we have seen before in Sect.~\ref{sec:big_data_for_NTMA}, a group of works applies big data techniques in the first phases of the KDD process. This approach allows the analytics phase to be performed using non-distributed frameworks (Fig.~\ref{fig:taxonomy-of-machine-learning-frameworks}).

For example, in~\cite{rathore_hadoop_2016} authors use Hadoop for real-time intrusion detection, but only computing feature values with MapReduce. Classic machine learning algorithms are used afterward on the reduced data. Similarly, Vassio et al.~\cite{vassio_users_2017} use big data approaches to reduce the data dimension, while the classification is done in a centralized manner, with traditional machine learning frameworks. Shibahara et al.~\cite{shibahara_malicious_2017} deploy a system to classify malicious URLs through neural networks, analyzing IP address hierarchies. Only the feature extraction is performed using the MapReduce paradigm.

In summary, we observe a majority of works adopting a single (big data) source, performing early data reduction with the approaches described in Sect.~\ref{sec:big_data_for_NTMA} (i.e., pre-processing data), and then performing machine learning analysis with traditional non-distributed platforms.

\subsection{Big data platforms enabling big NTMA}

A small group of papers performs the analytics process with big data approaches. Here different directions are taken. In \textit{Hashdoop}~\cite{fontugne_hashdoop_2014}, authors split the traffic into buckets according to a hash function applied to traffic features. Anomaly detection methods are then applied to each bucket, directly implemented as Map functions. Authors of~\cite{li_world_2016} present a distributed semi-supervised clustering technique on top of MapReduce, using a local spectral subspace approach to analyze YouTube user comment-based graphs. Authors of~\cite{lee_detecting_2011} perform both pre-processing of network data using MapReduce (e.g., to filter out non-HTTP GET packets from HTTP traffic logs) as well as simple analytics to summarize the network activity per client-server pairs. Lastly, the Tsinghua University Campus has tested in its network an entropy-based anomaly detection system that uses IPFIX flows and runs over Hadoop~\cite{tian_tadoop_2015}.

In traffic classification, Trevisan et al.~\cite{trevisan_awesome_2018} developed AWESoME, an SDN application that allows prioritizing traffic of critical Web services, while segregating others, even if they are running on the same cloud or served by the same CDN. To classify flows, the training, performed on large datasets, is implemented in Spark.

Considering other applications, some works consider the analysis of large amounts of non-traffic data with big data approaches. Authors in \cite{spina_snooping_2015} use MapReduce as a basis for a distributed crawler, which is applied to analyze over 300 million pages from Wikipedia to identify reliable editors, and subsequently detect editors that are likely vandals. Comarela et al.~\cite{comarela_studying_2013} focus on routing and implement a MapReduce algorithm to study multi-hop routing table distances. This function, applied over a TB of data, produces a measure of the variation of paths in different timescales.

In a nutshell, here big data platforms are enablers to scale the analysis on large datasets.

\subsection{The rare cases of online analysis}

Only a few works focus on online analysis and even fewer leverage big data techniques. Besides the previously cited~\cite{tian_tadoop_2015,rathore_hadoop_2016}, Apiletti et al. in~\cite{apiletti_selina_2016} developed SeLINA, a network analyzer that offers human-readable models of traffic data, combining unsupervised and supervised approaches for traffic inspection.
A specific framework for distributed network analytics that operates using Netflow and IPFIX flows is presented in~\cite{clemm_dna_2015}. Here, SDN controllers are used for the processing to improve scalability and analytics quality by dynamically adjusting traffic record generation.

In sum, online analysis in NTMA is mostly restricted to the techniques to perform high-speed traffic capture and processing, described in Sect.~\ref{sec:big_data_for_NTMA}. When it comes to big data analysis, NTMA researchers have mostly focused on batch analysis, thus not facing challenges of running algorithms on big data streaming.

\subsection{Takeaway}

The usage of ML seems to be widespread, especially for network security and anomaly detection. However, just some works use machine learning coupled with big data platforms. A general challenge when considering machine learning for big data analytics is indeed parallelization, which is not always easy to reach. Not all machine learning algorithms can be directly ported into a distributed framework, basically due to their inherent centralized designs. This hinders a wider adoption of big data platforms for the analytics stage, constraining works to perform data reduction at pre-processing stages.

In sum, in terms of complexity, most ML algorithms scale poorly with large datasets. When applied to the often humongous scale of NTMA data, they clearly cannot scale to typical NTMA scenarios.

\section{Challenges, open issues and ongoing work}~\label{sec:discussion}

We discuss some open issues and future directions we have identified after literature review.

\vspace{2mm}
\noindent
\textbf{1 --- Lack of a standard and context-generic big \ac{NTMA} platform: }
The data collection phase poses the major challenges for NTMA. The data transmission rate in computer networks keeps increasing, challenging the way probes collect and manage information. This is critical for probes that have to capture and process information on-the-fly and transmit results to centralized repositories. Flow-based approaches scale better, at the cost of losing details.

To solve the lack of flexible storage formats we have seen in Sect.~\ref{sub_sec:ntma_pre-processing} that researchers have developed APIs or layers that transform the data in pre-defined shapes. Those APIs are not generic and not comprehensive. As an example, in NTMA one would like to associate to a given IP address both its geographical (e.g., country) and logical (e.g., Autonomous System Number) location. There is no standard library that supports even these basic operations in the frameworks.

Considering analytics, few researchers tackled the problem from a big data perspective. There is a lack of generic approaches to access the data features, with the ability to run multiple analytics in a scalable way. Thus, researchers usually rely on single data sources and sequential/centralized algorithms, that are applied to reduced data (see Sect.~\ref{sub_sec:analisys_seq}).

In a nutshell, the community has yet to arrive at a generic platform for the big NTMA problem, and most solutions appear to be customized to solve a specific problems.

\vspace{2mm}
\noindent
\textbf{2 --- Lack of distributed machine learning and data mining algorithms in big data platforms limits NTMA:}
Several researchers started adopting machine learning solutions to tackle NTMA problems. However, as analyzed in Sect.~\ref{sec:big_data_analytics_for_NTMA}, most recent papers focus on ``small data'', with few of them addressing the scalability problem of typical big data cases. Most papers use big data techniques just in the first steps of the work, for data processing and feature extraction. Most of the machine learning analysis is then executed in a centralized fashion. This design represents a lack of opportunity. For example, applying machine learning with large datasets could produce more accurate models for NTMA applications.

From a scientific point of view, it is interesting to conjecture the causes of this gap: The reasons may be several, from the lack of expertise to platform limitations. We observe that the availability of machine learning algorithms in big data platforms is still at an early stage. Despite the availability of solutions like the Spark MLlib and ML tools that have started to provide some big data-tailored machine learning, not all algorithms are ported. Some of these algorithms are also simply hard to parallelize. Parallelization of traditional algorithms is a general problem that has to be faced for big data in general, and for big NTMA in particular.

\vspace{2mm}
\noindent
\textbf{3 --- Areas where big data NTMA are still missing:}
From Tab.~\ref{tab:bd_NTMA_summary}, it is easy to notice the lack of proposals in some important categories. For example, even though fault management is a category in which usually a great amount of data must be handled, few papers faced this problem with big data approaches. The reasons may be linked to what we discussed earlier, i.e., the lack of generic and standard NTMA platforms. Similarly, as examined in Sect.~\ref{sec:intro_tma}, some categories of NTMA applications (e.g., QoS/QoE management) are hardly faced with big data approaches.

\vspace{2mm}
\noindent
\textbf{4 --- Lack of relevant and/or public datasets limits reproducibility:}
To the extent of our survey, only two contributions disclose a public dataset, namely \cite{putina_telemetry-based_2018} and \cite{trevisan_awesome_2018}. Few works use open data, like the well-known MAWI dataset which is used for example in~\cite{vanerio_ensemble-learning_2017,fontugne_hashdoop_2014}, the (outdated) KDD CUP '99~\cite{frishman_cluster-based_2017,rathore_hadoop_2016}, and Kyoto2006~\cite{huang_new_2016}. Apart from these cases, public datasets are scarce and often not updated, posing limitations in reproducibility of researches as well as limiting the benchmark of new, possibly more scalable, solutions.

\vspace{2mm}
\noindent
\textbf{5 --- Ongoing projects on big NTMA:}
We have seen a solid increase in the adoption of big data approaches in NTMA. Yet, we observe a fragmented picture, with some limitations especially regarding interoperability and standardization. In fact, ad-hoc methodologies are proliferating, with no platform to support the community.

In this direction, Apache Spot was a promising platform (see Sect.~\ref{sub_sub_sec:big_data_acquisition}). Unfortunately, its development has stopped, thus questioning its practical adoption by the community and practitioners. PNDA is instead actively developed, and the project starts collecting interest from the community, albeit in its early stage. Beam\footnote{\url{https://beam.apache.org}} is a framework offering the unification of batch and streaming models, increasing portability and easing the work of programmers that do not need to write two code bases; yet, no applications for NTMA exists.

In sum, there is a lot of work to be done to arrive at a practical big data solution for NTMA applications. The NTMA community shall start creating synergies and consolidating solutions while relaying on the consolidated platforms offered by the big data community.

% \nocite{*}
\bibliographystyle{abbrv}
\bibliography{Big-Data}

\begin{thebibliography}{100}

\bibitem{abt_performance_2013}
S.~Abt, S.~Wener, and H.~Baier.
\newblock Performance {Evaluation} of {Classification} and {Feature}
  {Selection} {Algorithms} for {Netflow}-{Based} {Protocol} {Recognition}.
\newblock Proc. of the {INFORMATIK}, pages 2184--2197, 2013.

\bibitem{byungik_ahn_neuron_2012}
J.~B. Ahn.
\newblock Neuron {Machine}: {Parallel} and {Pipelined} {Digital}
  {Neurocomputing} {Architecture}.
\newblock Proc. of the {CyberneticsCom}, pages 143--147, 2012.

\bibitem{amiri_mutual_2011}
F.~Amiri, M.~R. Yousefi, C.~Lucas, A.~Shakery, and N.~Yazdani.
\newblock Mutual {Information}-{Based} {Feature} {Selection} for {Intrusion}
  {Detection} {Systems}.
\newblock {\em J. Netw. Comput. Appl.}, 34(4):1184--1199, 2011.

\bibitem{apiletti_selina_2016}
D.~Apiletti, E.~Baralis, T.~Cerquitelli, P.~Garza, D.~Giordano, M.~Mellia, and
  L.~Venturini.
\newblock {SeLINA}: {A} {Self}-{Learning} {Insightful} {Network} {Analyzer}.
\newblock {\em IEEE Trans. Netw. Service Manag.}, 13(3):696--710, 2016.

\bibitem{arzani_taking_2016}
B.~Arzani, S.~Ciraci, B.~T. Loo, A.~Schuster, and G.~Outhred.
\newblock Taking the {Blame} {Game} out of {Data} {Centers} {Operations} with
  {NetPoirot}.
\newblock Proc. of the {SIGCOMM}, pages 440--453, 2016.

\bibitem{bajaber_big_2016}
F.~Bajaber, R.~Elshawi, O.~Batarfi, A.~Altalhi, A.~Barnawi, and S.~Sakr.
\newblock Big {Data} 2.0 {Processing} {Systems}: {Taxonomy} and {Open}
  {Challenges}.
\newblock {\em J. Grid Comput.}, 14(3):379--405, 2016.

\bibitem{bar_large-scale_2014}
A.~Bar, A.~Finamore, P.~Casas, L.~Golab, and M.~Mellia.
\newblock Large-{Scale} {Network} {Traffic} {Monitoring} with {DBStream}, a
  {System} for {Rolling} {Big} {Data} {Analysis}.
\newblock Proc. of the {BigData}, pages 165--170, 2014.

\bibitem{beaver_finding_2010}
D.~Beaver, S.~Kumar, H.~C. Li, J.~Sobel, and P.~Vajgel.
\newblock Finding a {Needle} in {Haystack}: {Facebook}'s {Photo} {Storage}.
\newblock Proc. of the {OSDI}, pages 47--60, 2010.

\bibitem{bengio_learning_2009}
Y.~Bengio.
\newblock Learning {Deep} {Architectures} for {AI}.
\newblock {\em Foundations and Trends in Machine Learning}, 2(1):1--127, 2009.

\bibitem{bennett_numerically_2009}
J.~Bennett, R.~Grout, P.~Pebay, D.~Roe, and D.~Thompson.
\newblock Numerically {Stable}, {Single}-pass, {Parallel} {Statistics}
  {Algorithms}.
\newblock Proc. of the {CLUSTR}, pages 1--8, 2009.

\bibitem{benzidane_toward_2016}
K.~Benzidane, H.~E. Alloussi, O.~E. Warrak, L.~Fetjah, S.~J. Andaloussi, and
  A.~Sekkaki.
\newblock Toward a {Cloud}-{Based} {Security} {Intelligence} with {Big} {Data}
  {Processing}.
\newblock Proc. of the {NOMS}, pages 1089--1092, 2016.

\bibitem{bhuyan_network_2014}
M.~H. Bhuyan, D.~K. Bhattacharyya, and J.~K. Kalita.
\newblock Network {Anomaly} {Detection}: {Methods}, {Systems} and {Tools}.
\newblock {\em Commun. Surveys Tuts.}, 16(1):303--336, 2014.

\bibitem{bifet_big_2014}
A.~Bifet and G.~D.~F. Morales.
\newblock Big {Data} {Stream} {Learning} with {SAMOA}.
\newblock Proc. of the {ICDMW}, pages 1199--1202, 2014.

\bibitem{big_data_working_group_big_2014}
{Big Data Working Group}.
\newblock Big {Data} {Taxonomy}.
\newblock Technical report, Cloud Security Alliance, 2014.

\bibitem{boutaba_comprehensive_2018}
R.~Boutaba, M.~A. Salahuddin, N.~Limam, S.~Ayoubi, N.~Shahriar,
  F.~Estrada-Solano, and O.~M. Caicedo.
\newblock A {Comprehensive} {Survey} on {Machine} {Learning} for {Networking}:
  {Evolution}, {Applications} and {Research} {Opportunities}.
\newblock {\em J. Internet Serv. Appl.}, 9(16), 2018.

\bibitem{bujlow_independent_2015}
T.~Bujlow, V.~Carela-Espa{\~n}ol, and P.~Barlet-Ros.
\newblock Independent {Comparison} of {Popular} {DPI} {Tools} for {Traffic}
  {Classification}.
\newblock {\em Comput. Netw.}, 76(C):75--89, 2015.

\bibitem{callado_survey_2009}
A.~Callado, C.~Kamienski, G.~Szabo, B.~P. Gero, J.~Kelner, S.~Fernandes, and
  D.~Sadok.
\newblock A {Survey} on {Internet} {Traffic} {Identification}.
\newblock {\em Commun. Surveys Tuts.}, 11(3):37--52, 2009.

\bibitem{casas_gml_2017}
P.~Casas, J.~Vanerio, and K.~Fukuda.
\newblock {GML} {Learning}, a {Generic} {Machine} {Learning} {Model} for
  {Network} {Measurements} {Analysis}.
\newblock Proc. of the {CNSM}, pages 1--9, 2017.

\bibitem{cattell_scalable_2011}
R.~Cattell.
\newblock Scalable {SQL} and {NoSQL} {Data} {Stores}.
\newblock {\em SIGMOD Rec.}, 39(4):12--27, 2011.

\bibitem{cermak_real-time_2016}
M.~{\v C}erm{\'a}k, T.~Jirs{\'i}k, and M.~La{\v s}tovi{\v c}ka.
\newblock Real-{Time} {Analysis} of {Netflow} {Data} for {Generating} {Network}
  {Traffic} {Statistics} {Using} {Apache} {Spark}.
\newblock Proc. of the {NOMS}, pages 1019--1020, 2016.

\bibitem{chaiken_scope_2008}
R.~Chaiken, B.~Jenkins, P.-A. Larson, B.~Ramsey, D.~Shakib, S.~Weaver, and
  J.~Zhou.
\newblock {SCOPE}: {Easy} and {Efficient} {Parallel} {Processing} of {Massive}
  {Data} {Sets}.
\newblock {\em Proc. VLDB Endow.}, 1(2):1265--1276, 2008.

\bibitem{chandramouli_model-driven_2017}
M.~Chandramouli and A.~Clemm.
\newblock Model-{Driven} {Analytics} in {SDN} {Networks}.
\newblock Proc. of the {IM}, pages 668--673, 2017.

\bibitem{chang_bigtable_2008}
F.~Chang, J.~Dean, S.~Ghemawat, W.~C. Hsieh, D.~A. Wallach, M.~Burrows,
  T.~Chandra, A.~Fikes, and R.~E. Gruber.
\newblock Bigtable: {A} {Distributed} {Storage} {System} for {Structured}
  {Data}.
\newblock {\em ACM Trans. Comput. Syst.}, 26(2):4:1--4:26, 2008.

\bibitem{chen_big_2014}
M.~Chen, S.~Mao, and Y.~Liu.
\newblock Big {Data}: {A} {Survey}.
\newblock {\em Mobile Netw. Appl.}, 19(2):171--209, 2014.

\bibitem{chen_survey_2006}
Y.~Chen, Y.~Li, X.-Q. Cheng, and L.~Guo.
\newblock Survey and {Taxonomy} of {Feature} {Selection} {Algorithms} in
  {Intrusion} {Detection} {System}.
\newblock Proc. of the Inscrypt, pages 153--167, 2006.

\bibitem{ciresan_multi-column_2012}
D.~Ciresan, U.~Meier, and J.~Schmidhuber.
\newblock Multi-{Column} {Deep} {Neural} {Networks} for {Image}
  {Classification}.
\newblock Proc. of the {CVPR}, pages 3642--3649, 2012.

\bibitem{clemm_dna_2015}
A.~Clemm, M.~Chandramouli, and S.~Krishnamurthy.
\newblock {DNA}: {An} {SDN} {Framework} for {Distributed} {Network}
  {Analytics}.
\newblock Proc. of the {IM}, pages 9--17, 2015.

\bibitem{cogranne_detecting_2018}
R.~Cogranne, G.~Doyen, N.~Ghadban, and B.~Hammi.
\newblock Detecting {Botclouds} at {Large} {Scale}: {A} {Decentralized} and
  {Robust} {Detection} {Method} for {Multi}-{Tenant} {Virtualized}
  {Environments}.
\newblock {\em IEEE Trans. Netw. Service Manag.}, 15(1):68--82, 2018.

\bibitem{comarela_studying_2013}
G.~Comarela, G.~G{\"u}rsun, and M.~Crovella.
\newblock Studying {Interdomain} {Routing} over {Long} {Timescales}.
\newblock Proc. of the {IMC}, pages 227--234, 2013.

\bibitem{decandia_dynamo_2007}
G.~DeCandia, D.~Hastorun, M.~Jampani, G.~Kakulapati, A.~Lakshman, A.~Pilchin,
  S.~Sivasubramanian, P.~Vosshall, and W.~Vogels.
\newblock Dynamo: {Amazon}'s {Highly} {Available} {Key}-{Value} {Store}.
\newblock {\em SIGOPS Oper. Syst. Rev.}, 41(6):205--220, 2007.

\bibitem{deri_10_2013}
L.~Deri, A.~Cardigliano, and F.~Fusco.
\newblock 10 {Gbit} {Line} {Rate} {Packet}-to-{Disk} {Using} n2disk.
\newblock Proc. of the {TMA}, pages 441--446, 2013.

\bibitem{sakr_big_2018}
I.~Drago, M.~Mellia, and A.~D{\textquoteright}Alconzo.
\newblock Big {Data} in {Computer} {Network} {Monitoring}.
\newblock In {\em Encyclopedia of {Big} {Data} {Technologies}}. Springer
  International Publishing, 2018.

\bibitem{erman_traffic_2006}
J.~Erman, M.~Arlitt, and A.~Mahanti.
\newblock Traffic {Classification} {Using} {Clustering} {Algorithms}.
\newblock Proc. of the {MineNet}, pages 281--286, 2006.

\bibitem{fahad_survey_2014}
A.~Fahad, N.~Alshatri, Z.~Tari, A.~Alamri, I.~Khalil, A.~Y. Zomaya, S.~Foufou,
  and A.~Bouras.
\newblock A {Survey} of {Clustering} {Algorithms} for {Big} {Data}: {Taxonomy}
  and {Empirical} {Analysis}.
\newblock {\em IEEE Trans. Emerg. Topics Comput.}, 2(3):267--279, 2014.

\bibitem{fayyad_data_1996}
U.~M. Fayyad, G.~Piatetsky-Shapiro, and P.~Smyth.
\newblock From {Data} {Mining} to {Knowledge} {Discovery} in {Databases}.
\newblock {\em AI Magazine}, 17(3):37--54, 1996.

\bibitem{fiadino_call_2017}
P.~Fiadino, V.~Ponce-Lopez, J.~Antonio, M.~Torrent-Moreno, and A.~D'Alconzo.
\newblock Call {Detail} {Records} for {Human} {Mobility} {Studies}: {Taking}
  {Stock} of the {Situation} in the "{Always} {Connected} {Era}".
\newblock Proc. of the Big-{DAMA}, pages 43--48, 2017.

\bibitem{fiadino_grasping_2016}
S.~Fiadino, P.~Casas, A.~D{\textquoteright}Alconzo, M.~Schiavone, and A.~Baer.
\newblock Grasping {Popular} {Applications} in {Cellular} {Networks} with {Big}
  {Data} {Analytics} {Platforms}.
\newblock {\em IEEE Trans. Netw. Service Manag.}, 13(3):681--695, 2016.

\bibitem{fontugne_hashdoop_2014}
R.~Fontugne, J.~Mazel, and K.~Fukuda.
\newblock Hashdoop: {A} {Mapreduce} {Framework} for {Network} {Anomaly}
  {Detection}.
\newblock Proc. of the {INFOCOM} {WKSHPS}, pages 494--499, 2014.

\bibitem{frishman_cluster-based_2017}
G.~Frishman, Y.~Ben-Itzhak, and O.~Margalit.
\newblock Cluster-{Based} {Load} {Balancing} for {Better} {Network} {Security}.
\newblock Proc. of the Big-{DAMA}, pages 7--12, 2017.

\bibitem{fuchs_implications_2012}
C.~Fuchs.
\newblock Implications of {Deep} {Packet} {Inspection} ({DPI}) {Internet}
  {Surveillance} for {Society}.
\newblock Technical Report~1, Uppsala University, Media and Communication
  Studies, 2012.

\bibitem{fusco_pcapindex_2012}
F.~Fusco, X.~Dimitropoulos, M.~Vlachos, and L.~Deri.
\newblock {pcapIndex}: {An} {Index} for {Network} {Packet} {Traces} with
  {Legacy} {Compatibility}.
\newblock {\em SIGCOMM Comput. Commun. Rev.}, 42(1):47--53, 2012.

\bibitem{fusco_net-fli_2010}
F.~Fusco, M.~P. Stoecklin, and M.~Vlachos.
\newblock {NET}-{FLi}: {On}-the-{Fly} {Compression}, {Archiving} and {Indexing}
  of {Streaming} {Network} {Traffic}.
\newblock {\em Proc. VLDB Endow.}, 3(1-2):1382--1393, 2010.

\bibitem{garcia_efficient_2018}
J.~Garcia and T.~Korhonen.
\newblock Efficient {Distribution}-{Derived} {Features} for {High}-{Speed}
  {Encrypted} {Flow} {Classification}.
\newblock Proc. of the {NetAI}, pages 21--27, 2018.

\bibitem{garcia-teodoro_anomaly-based_2009}
P.~Garc{\'i}a-Teodoro, J.~D{\'i}az-Verdejo, G.~Maci{\'a}-Fern{\'a}ndez, and
  E.~V{\'a}zquez.
\newblock Anomaly-{Based} {Network} {Intrusion} {Detection}: {Techniques},
  {Systems} and {Challenges}.
\newblock {\em Comput. Secur.}, 28:18--28, 2009.

\bibitem{gates_building_2009}
A.~F. Gates, O.~Natkovich, S.~Chopra, P.~Kamath, S.~M. Narayanamurthy,
  C.~Olston, B.~Reed, S.~Srinivasan, and U.~Srivastava.
\newblock Building a {High}-{Level} {Dataflow} {System} on {Top} of
  {Map}-{Reduce}: {The} {Pig} {Experience}.
\newblock {\em Proc. VLDB Endow.}, 2(2):1414--1425, 2009.

\bibitem{ghemawat_google_2003}
S.~Ghemawat, H.~Gobioff, and S.-T. Leung.
\newblock The {Google} {File} {System}.
\newblock {\em SIGOPS Oper. Syst. Rev.}, 37(5):29--43, 2003.

\bibitem{gonzalez_net2vec_2017}
R.~Gonzalez, F.~Manco, A.~Garcia-Duran, J.~Mendes, F.~Huici, S.~Niccolini, and
  M.~Niepert.
\newblock Net2vec: {Deep} {Learning} for the {Network}.
\newblock Proc. of the Big-{DAMA}, pages 13--18, 2017.

\bibitem{hameed_efficacy_2016}
S.~Hameed and U.~Ali.
\newblock Efficacy of {Live} {DDoS} {Detection} with {Hadoop}.
\newblock Proc. of the {NOMS}, pages 488--494, 2016.

\bibitem{han_survey_2011}
J.~Han, H.~E, G.~Le, and J.~Du.
\newblock Survey on {NoSQL} {Database}.
\newblock Proc. of the {ICPCA}, pages 363--366, 2011.

\bibitem{harper_cookbook_2018}
R.~Harper and P.~Tee.
\newblock Cookbook, a {Recipe} for {Fault} {Localization}.
\newblock Proc. of the {NOMS}, pages 1--6, 2018.

\bibitem{hastie_elements_2009}
T.~Hastie, R.~Tibshirani, and J.~Friedman.
\newblock {\em The {Elements} of {Statistical} {Learning}: {Data} {Mining},
  {Inference} and {Prediction}}.
\newblock Springer-Verlag New York, 2 edition, 2009.

\bibitem{hofstede_flow_2014}
R.~Hofstede, P.~{\v C}eleda, B.~Trammell, I.~Drago, R.~Sadre, A.~Sperotto, and
  A.~Pras.
\newblock Flow {Monitoring} {Explained}: {From} {Packet} {Capture} to {Data}
  {Analysis} with {NetFlow} and {IPFIX}.
\newblock {\em Commun. Surveys Tuts.}, 16(4):2037--2064, 2014.

\bibitem{hofstede_ssh_2014}
R.~Hofstede, L.~Hendriks, A.~Sperotto, and A.~Pras.
\newblock {SSH} {Compromise} {Detection} {Using} {NetFlow}/{IPFIX}.
\newblock {\em SIGCOMM Comput. Commun. Rev.}, 44(5):20--26, 2014.

\bibitem{hu_toward_2014}
H.~Hu, Y.~Wen, T.-S. Chua, and X.~Li.
\newblock Toward {Scalable} {Systems} for {Big} {Data} {Analytics}: {A}
  {Technology} {Tutorial}.
\newblock {\em IEEE Access}, 2:652--687, 2014.

\bibitem{huang_new_2016}
T.~Huang, H.~Sethu, and N.~Kandasamy.
\newblock A {New} {Approach} to {Dimensionality} {Reduction} for {Anomaly}
  {Detection} in {Data} {Traffic}.
\newblock {\em IEEE Trans. Netw. Service Manag.}, 13(3):651--665, 2016.

\bibitem{ibrahim_study_2015}
L.~T. Ibrahim, R.~Hassan, K.~Ahmad, and A.~N. Asat.
\newblock A {Study} on {Improvement} of {Internet} {Traffic} {Measurement} and
  {Analysis} using {Hadoop} {System}.
\newblock Proc. of the {ICEEI}, pages 462--466, 2015.

\bibitem{iglesias_analysis_2015}
F.~Iglesias and T.~Zseby.
\newblock Analysis of {Network} {Traffic} {Features} for {Anomaly} {Detection}.
\newblock {\em Mach. Learn.}, 101(1-3):59--84, 2015.

\bibitem{intel_data_2014}
{Intel}.
\newblock Data {Plane} {Development} {Kit} ({DPDK}), 2014.

\bibitem{jiang_cross-domain_2008}
W.~Jiang, E.~Zavesky, S.-F. Chang, and A.~Loui.
\newblock Cross-{Domain} {Learning} {Methods} for {High}-{Level} {Visual}
  {Concept} {Classification}.
\newblock Proc. of the {ICIP}, pages 161--164, 2008.

\bibitem{karagiannis_blinc_2005}
T.~Karagiannis, K.~Papagiannaki, and M.~Faloutsos.
\newblock {BLINC}: {Multilevel} {Traffic} {Classification} in the {Dark}.
\newblock Proc. of the {SIGCOMM}, pages 229--240, 2005.

\bibitem{kasai_network_2016}
H.~Kasai, W.~Kellerer, and M.~Kleinsteuber.
\newblock Network {Volume} {Anomaly} {Detection} and {Identification} in
  {Large}-{Scale} {Networks} {Based} on {Online} {Time}-{Structured} {Traffic}
  {Tensor} {Tracking}.
\newblock {\em IEEE Trans. Netw. Service Manag.}, 13(3):636--650, 2016.

\bibitem{kim_combined_2004}
M.~Kim, H.~Na, K.~Chae, H.~Bang, and J.~Na.
\newblock A {Combined} {Data} {Mining} {Approach} for {DDoS} {Attack}
  {Detection}.
\newblock Proc. of the {ICOIN}, 2004.

\bibitem{kobayashi_mining_2018}
S.~Kobayashi, K.~Otomo, K.~Fukuda, and H.~Esaki.
\newblock Mining {Causality} of {Network} {Events} in {Log} {Data}.
\newblock {\em IEEE Trans. Netw. Service Manag.}, 15(1):53--67, 2018.

\bibitem{lakshman_cassandra_2010}
A.~Lakshman and P.~Malik.
\newblock Cassandra: {A} {Decentralized} {Structured} {Storage} {System}.
\newblock {\em SIGOPS Oper. Syst. Rev.}, 44(2):35--40, 2010.

\bibitem{laney_3d_2001}
D.~Laney.
\newblock 3d {Data} {Management}: {Controlling} {Data} {Volume}, {Velocity},
  and {Variety}.
\newblock Technical report, META Group, 2001.

\bibitem{le_building_2013}
Q.~V. Le.
\newblock Building {High}-{Level} {Features} using {Large} {Scale}
  {Unsupervised} {Learning}.
\newblock Proc. of the {ICASSP}, pages 8595--8598, 2013.

\bibitem{lee_internet_2010}
Y.~Lee, W.~Kang, and H.~Son.
\newblock An {Internet} {Traffic} {Analysis} {Method} with {MapReduce}.
\newblock Proc. of the {NOMS}, pages 357--361, 2010.

\bibitem{lee_detecting_2011}
Y.~Lee and Y.~Lee.
\newblock Detecting {DDoS} {Attacks} with {Hadoop}.
\newblock Proc. of the {CoNEXT}, pages 7:1--7:2, 2011.

\bibitem{lee_toward_2013}
Y.~Lee and Y.~Lee.
\newblock Toward {Scalable} {Internet} {Traffic} {Measurement} and {Analysis}
  with {Hadoop}.
\newblock {\em SIGCOMM Comput. Commun. Rev.}, 43(1):5--13, 2013.

\bibitem{li_supervised_2013}
B.~Li, M.~H. Gunes, G.~Bebis, and J.~Springer.
\newblock A {Supervised} {Machine} {Learning} {Approach} to {Classify} {Host}
  {Roles} on {Line} {Using} {sFlow}.
\newblock Proc. of the {HPPN}, pages 53--60, 2013.

\bibitem{li_deep_2018}
M.~Li, C.~Lumezanu, B.~Zong, and H.~Chen.
\newblock Deep {Learning} {IP} {Network} {Representations}.
\newblock Proc. of the Big-{DAMA}, pages 33--39, 2018.

\bibitem{li_world_2016}
Y.~Li, O.~Martinez, X.~Chen, Y.~Li, and J.~E. Hopcroft.
\newblock In a {World} {That} {Counts}: {Clustering} and {Detecting} {Fake}
  {Social} {Engagement} at {Scale}.
\newblock Proc. of the {WWW}, pages 111--120, 2016.

\bibitem{li_building_2009}
Y.~Li, J.-L. Wang, Z.-H. Tian, T.-B. Lu, and C.~Young.
\newblock Building {Lightweight} {Intrusion} {Detection} {System} {Using}
  {Wrapper}-{Based} {Feature} {Selection} {Mechanisms}.
\newblock {\em Comput. Secur.}, 28(6):466--475, 2009.

\bibitem{liao_intrusion_2013}
H.-J. Liao, C.-H.~R. Lin, Y.-C. Lin, and K.-Y. Tung.
\newblock Intrusion {Detection} {System}: {A} {Comprehensive} {Review}.
\newblock {\em J. Netw. Comput. Appl.}, 36(1):16--24, 2013.

\bibitem{maier_enriching_2008}
G.~Maier, R.~Sommer, H.~Dreger, A.~Feldmann, V.~Paxson, and F.~Schneider.
\newblock Enriching {Network} {Security} {Analysis} with {Time} {Travel}.
\newblock Proc. of the {SIGCOMM}, pages 183--194, 2008.

\bibitem{manyika_big_2011}
J.~Manyika and {others}.
\newblock Big {Data}: {The} {Next} {Frontier} for {Innovation}, {Competition},
  and {Productivity}, 2011.

\bibitem{marchal_big_2014}
S.~Marchal, X.~Jiang, R.~State, and T.~Engel.
\newblock A {Big} {Data} {Architecture} for {Large} {Scale} {Security}
  {Monitoring}.
\newblock Proc. of the {BigData}.{Congress}, pages 56--63, 2014.

\bibitem{mchugh_testing_2000}
J.~McHugh.
\newblock Testing {Intrusion} {Detection} {Systems}: {A} {Critique} of the 1998
  and 1999 {Darpa} {Intrusion} {Detection} {System} {Evaluations} as
  {Performed} by {Lincoln} {Laboratory}.
\newblock {\em ACM Trans. Inf. Syst. Secur.}, 3(4):262--294, 2000.

\bibitem{mckusick_gfs_2009}
M.~K. McKusick and S.~Quinlan.
\newblock {GFS}: {Evolution} on {Fast}-forward.
\newblock {\em Queue}, 7(7):10:10--10:20, 2009.

\bibitem{mikolov_strategies_2011}
T.~Mikolov, A.~Deoras, D.~Povey, L.~Burget, and J.~Cernocky.
\newblock Strategies for {Training} {Large} {Scale} {Neural} {Network}
  {Language} {Models}.
\newblock Proc. of the {ASRU}, pages 196--201, 2011.

\bibitem{mit_lincoln_laboratories_darpa_1999}
{"MIT Lincoln Laboratories"}.
\newblock {DARPA} {Intrusion} {Detection} {Data} {Sets}, 1999.

\bibitem{mulinka_stream-based_2018}
P.~Mulinka and P.~Casas.
\newblock Stream-{Based} {Machine} {Learning} for {Network} {Security} and
  {Anomaly} {Detection}.
\newblock Proc. of the Big-{DAMA}, pages 1--7, 2018.

\bibitem{murgia_lightweight_2016}
A.~Murgia, G.~Ghidini, S.~P. Emmons, and P.~Bellavista.
\newblock Lightweight {Internet} {Traffic} {Classification}: {A}
  {Subject}-{Based} {Solution} with {Word} {Embeddings}.
\newblock Proc. of the {SMARTCOMP}, pages 1--8, 2016.

\bibitem{nagele_large-scale_2011}
W.~Nagele.
\newblock Large-{Scale} {PCAP} {Data} {Analysis} {Using} {Apache} {Hadoop},
  2011.

\bibitem{nguyen_survey_2008}
T.~T. Nguyen and G.~Armitage.
\newblock A {Survey} of {Techniques} for {Internet} {Traffic} {Classification}
  {Using} {Machine} {Learning}.
\newblock {\em Commun. Surveys Tuts.}, 10(4):56--76, 2008.

\bibitem{nickless_combining_2000}
B.~Nickless.
\newblock Combining {Cisco} {NetFlow} {Exports} with {Relational} {Database}
  {Technology} for {Usage} {Statistics}, {Intrusion} {Detection}, and {Network}
  {Forensics}.
\newblock Proc. of the {LISA}, pages 285--290, 2000.

\bibitem{orsini_bgpstream_2016}
C.~Orsini, A.~King, D.~Giordano, V.~Giotsas, and A.~Dainotti.
\newblock {BGPStream}: {A} {Software} {Framework} for {Live} and {Historical}
  {BGP} {Data} {Analysis}.
\newblock Proc. of the {IMC}, pages 429--444, 2016.

\bibitem{otomo_finding_2018}
K.~Otomo, S.~Kobayashi, K.~Fukuda, and H.~Esaki.
\newblock Finding {Anomalies} in {Network} {System} {Logs} with {Latent}
  {Variables}.
\newblock Proc. of the Big-{DAMA}, pages 8--14, 2018.

\bibitem{owen_mahout_2011}
S.~Owen, R.~Anil, T.~Dunning, and E.~Friedman.
\newblock {\em Mahout in {Action}}.
\newblock Manning Publications Co., Greenwich, CT, USA, 2011.

\bibitem{patterson_netflow_2013}
M.~Patterson.
\newblock {NetFlow} {Vs}. {Packet} {Analysis}, 2013.

\bibitem{pebay_design_2011}
P.~Pebay, D.~Thompson, J.~Bennett, and A.~Mascarenhas.
\newblock Design and {Performance} of a {Scalable}, {Parallel} {Statistics}
  {Toolkit}.
\newblock Proc. of the {IPDPS}, pages 1475--1484, 2011.

\bibitem{pike_interpreting_2005}
R.~Pike, S.~Dorward, R.~Griesemer, and S.~Quinlan.
\newblock Interpreting the {Data}: {Parallel} {Analysis} with {Sawzall}.
\newblock {\em Sci. Programming}, 13(4):277--298, 2005.

\bibitem{putina_telemetry-based_2018}
A.~Putina, D.~Rossi, A.~Bifet, S.~Barth, D.~Pletcher, C.~Precup, and
  P.~Nivaggioli.
\newblock Telemetry-{Based} {Stream}-{Learning} of {BGP} {Anomalies}.
\newblock Proc. of the Big-{DAMA}, pages 15--20, 2018.

\bibitem{rathore_hadoop_2016}
M.~M. Rathore, A.~Paul, A.~Ahmad, S.~Rho, M.~Imran, and M.~Guizani.
\newblock Hadoop {Based} {Real}-{Time} {Intrusion} {Detection} for
  {High}-{Speed} {Networks}.
\newblock Proc. of the {GLOBECOM}, pages 1--6, 2016.

\bibitem{raykar_fast_2007}
V.~C. Raykar, R.~Duraiswami, and B.~Krishnapuram.
\newblock A {Fast} {Algorithm} for {Learning} {Large} {Scale} {Preference}
  {Relations}.
\newblock Proc. of the {AISTATS}, pages 388--395, 2007.

\bibitem{sacerdoti_wide_2003}
F.~D. Sacerdoti, M.~J. Katz, M.~L. Massie, and D.~E. Culler.
\newblock Wide {Area} {Cluster} {Monitoring} with {Ganglia}.
\newblock Proc. of the {CLUSTR}-03, pages 289--298, 2003.

\bibitem{sakr_family_2013}
S.~Sakr, A.~Liu, and A.~G. Fayoumi.
\newblock The {Family} of {Mapreduce} and {Large}-scale {Data} {Processing}
  {Systems}.
\newblock {\em ACM Comput. Surv.}, 46(1):1--44, 2013.

\bibitem{samak_scalable_2012}
T.~Samak, D.~Gunter, and V.~Hendrix.
\newblock Scalable {Analysis} of {Network} {Measurements} with {Hadoop} and
  {Pig}.
\newblock Proc. of the {NOMS}, pages 1254--1259, 2012.

\bibitem{santos_network_2015}
O.~Santos.
\newblock {\em Network {Security} with {NetFlow} and {IPFIX}: {Big} {Data}
  {Analytics} for {Information} {Security}}.
\newblock Cisco Press, Indianapolis, 1 edition, 2015.

\bibitem{sarlis_datix_2015}
D.~Sarlis, N.~Papailiou, I.~Konstantinou, G.~Smaragdakis, and N.~Koziris.
\newblock Datix: {A} {System} for {Scalable} {Network} {Analytics}.
\newblock {\em SIGCOMM Comput. Commun. Rev.}, 45(5):21--28, 2015.

\bibitem{schiff_netslicer_2018}
L.~Schiff, O.~Ziv, M.~Jaeger, and S.~Schmid.
\newblock {NetSlicer}: {Automated} and {Traffic}-{Pattern} {Based}
  {Application} {Clustering} in {Datacenters}.
\newblock Proc. of the Big-{DAMA}, pages 21--26, 2018.

\bibitem{shadi_hierarchical_2017}
K.~Shadi, P.~Natarajan, and C.~Dovrolis.
\newblock Hierarchical {IP} flow clustering.
\newblock Proc. of the Big-{DAMA}, pages 25--30, 2017.

\bibitem{sherry_blindbox_2015}
J.~Sherry, C.~Lan, R.~A. Popa, and S.~Ratnasamy.
\newblock {BlindBox}: {Deep} {Packet} {Inspection} over {Encrypted} {Traffic}.
\newblock Proc. of the {SIGCOMM}, pages 213--226, 2015.

\bibitem{shibahara_malicious_2017}
T.~Shibahara, K.~Yamanishi, Y.~Takata, D.~Chiba, M.~Akiyama, T.~Yagi,
  Y.~Ohsita, and M.~Murata.
\newblock Malicious {URL} {Sequence} {Detection} {Using} {Event} {De}-{Noising}
  {Convolutional} {Neural} {Network}.
\newblock Proc. of the {ICC}, pages 1--7, 2017.

\bibitem{sperotto_overview_2010}
A.~Sperotto, G.~Schaffrath, R.~Sadre, C.~Morariu, A.~Pras, and B.~Stiller.
\newblock An {Overview} of {IP} {Flow}-{Based} {Intrusion} {Detection}.
\newblock {\em Commun. Surveys Tuts.}, 12(3):343--356, 2010.

\bibitem{spina_snooping_2015}
M.~Spina, D.~Rossi, M.~Sozio, S.~Maniu, and B.~Cautis.
\newblock Snooping {Wikipedia} {Vandals} with {MapReduce}.
\newblock Proc. of the {ICC}, pages 1146--1151, 2015.

\bibitem{stonebraker_newsql_2011}
M.~Stonebraker.
\newblock {NewSQL}: {An} {Alternative} to {NoSQL} and {Old} {SQL} for {New}
  {OLTP} {Apps}, 2011.

\bibitem{sun_sparse_2010}
P.~Sun and X.~Yao.
\newblock Sparse {Approximation} {Through} {Boosting} for {Learning} {Large}
  {Scale} {Kernel} {Machines}.
\newblock {\em IEEE Trans. Neural Netw.}, 21(6):883--894, 2010.

\bibitem{tazaki_matatabi_2014}
H.~Tazaki, K.~Okada, Y.~Sekiya, and Y.~Kadobayashi.
\newblock {MATATABI}: {Multi}-{Layer} {Threat} {Analysis} {Platform} with
  {Hadoop}.
\newblock Proc. of the {BADGERS}, pages 75--82, 2014.

\bibitem{thusoo_hive_2009}
A.~Thusoo, J.~S. Sarma, N.~Jain, Z.~Shao, P.~Chakka, S.~Anthony, H.~Liu,
  P.~Wyckoff, and R.~Murthy.
\newblock Hive: {A} {Warehousing} {Solution} over a {Map}-{Reduce} {Framework}.
\newblock {\em Proc. VLDB Endow.}, 2(2):1626--1629, 2009.

\bibitem{tian_tadoop_2015}
G.~Tian, Z.~Wang, X.~Yin, Z.~Li, X.~Shi, Z.~Lu, C.~Zhou, Y.~Yu, and D.~Wu.
\newblock {TADOOP}: {Mining} {Network} {Traffic} {Anomalies} with {Hadoop}.
\newblock Proc. of the {SecureComm}, pages 175--192, 2015.

\bibitem{trammell_mplane_2014}
B.~Trammell, P.~Casas, D.~Rossi, A.~Bar, Z.~Houidi, I.~Leontiadis, T.~Szemethy,
  and M.~Mellia.
\newblock {mPlane}: {An} {Intelligent} {Measurement} {Plane} for the
  {Internet}.
\newblock {\em IEEE Commun. Mag.}, 52(5):148--156, 2014.

\bibitem{trevisan_towards_2016}
M.~Trevisan, I.~Drago, M.~Mellia, and M.~M. Munafo.
\newblock Towards {Web} {Service} {Classification} using {Addresses} and {DNS}.
\newblock Proc. of the {TRAC}, pages 38--43, 2016.

\bibitem{trevisan_awesome_2018}
M.~Trevisan, I.~Drago, M.~Mellia, H.~H. Song, and M.~Baldi.
\newblock {AWESoME}: {Big} {Data} for {Automatic} {Web} {Service} {Management}
  in {SDN}.
\newblock {\em IEEE Trans. Netw. Service Manag.}, 15(1):13--26, 2018.

\bibitem{trevisan_traffic_2017}
M.~Trevisan, A.~Finamore, M.~Mellia, M.~Munafo, and D.~Rossi.
\newblock Traffic {Analysis} with {Off}-the-{Shelf} {Hardware}: {Challenges}
  and {Lessons} {Learned}.
\newblock {\em IEEE Commun. Mag.}, 55(3):163--169, 2017.

\bibitem{tsai_big_2015}
C.-W. Tsai, C.-F. Lai, H.-C. Chao, and A.~V. Vasilakos.
\newblock Big {Data} {Analytics}: {A} {Survey}.
\newblock {\em J. Big Data}, 2(1):21, 2015.

\bibitem{uwagbole_applied_2017}
S.~O. Uwagbole, W.~J. Buchanan, and L.~Fan.
\newblock Applied {Machine} {Learning} {Predictive} {Analytics} to {SQL}
  {Injection} {Attack} {Detection} and {Prevention}.
\newblock Proc. of the {IM}, pages 1087--1090, 2017.

\bibitem{vaarandi_unsupervised_2018}
R.~Vaarandi, B.~Blumbergs, and M.~Kont.
\newblock An {Unsupervised} {Framework} for {Detecting} {Anomalous} {Messages}
  from {Syslog} {Log} {Files}.
\newblock Proc. of the {NOMS}, pages 1--6, 2018.

\bibitem{valenti_identifying_2011}
S.~Valenti and D.~Rossi.
\newblock Identifying {Key} {Features} for {P}2p {Traffic} {Classification}.
\newblock Proc. of the {ICC}, pages 1--6, 2011.

\bibitem{valenti_reviewing_2013}
S.~Valenti, D.~Rossi, A.~Dainotti, A.~Pescap{\`e}, A.~Finamore, and M.~Mellia.
\newblock Reviewing {Traffic} {Classification}.
\newblock In {\em Data {Traffic} {Monitoring} and {Analysis} - {From}
  {Measurement}, {Classification}, and {Anomaly} {Detection} to {Quality} of
  {Experience}}. Springer, Heidelberg, 1 edition, 2013.

\bibitem{vanerio_ensemble-learning_2017}
J.~Vanerio and P.~Casas.
\newblock Ensemble-{Learning} {Approaches} for {Network} {Security} and
  {Anomaly} {Detection}.
\newblock Proc. of the Big-{DAMA}, pages 1--6, 2017.

\bibitem{vassio_users_2017}
L.~Vassio, D.~Giordano, M.~Trevisan, M.~Mellia, and A.~P.~C. da~Silva.
\newblock Users' {Fingerprinting} {Techniques} from {TCP} {Traffic}.
\newblock Proc. of the Big-{DAMA}, pages 49--54, 2017.

\bibitem{wassermann_netperftrace_2017}
S.~Wassermann, P.~Casas, T.~Cuvelier, and B.~Donnet.
\newblock {NETPerfTrace}: {Predicting} {Internet} {Path} {Dynamics} and
  {Performance} with {Machine} {Learning}.
\newblock Proc. of the Big-{DAMA}, pages 31--36, 2017.

\bibitem{wullink_entrada_2016}
M.~Wullink, G.~C.~M. Moura, M.~M{\"u}ller, and C.~Hesselman.
\newblock {ENTRADA}: {A} {High}-{Performance} {Network} {Traffic} {Data}
  {Stream}.
\newblock Proc. of the {NOMS}, pages 913--918, 2016.

\bibitem{yuan_privacy_2014}
J.~Yuan and S.~Yu.
\newblock Privacy {Preserving} {Back}-{Propagation} {Neural} {Network}
  {Learning} {Made} {Practical} with {Cloud} {Computing}.
\newblock {\em IEEE Trans. Parallel Distrib. Syst.}, 25(1):212--221, 2014.

\bibitem{zhang_parallel_2016}
Y.~Zhang, T.~Cao, S.~Li, X.~Tian, L.~Yuan, H.~Jia, and A.~V. Vasilakos.
\newblock Parallel {Processing} {Systems} for {Big} {Data}: {A} {Survey}.
\newblock {\em Proc. IEEE}, 104(11):2114--2136, 2016.

\bibitem{zhao_data_2018}
J.~Zhao, T.~Tiplea, R.~Mortier, J.~Crowcroft, and L.~Wang.
\newblock Data {Analytics} {Service} {Composition} and {Deployment} on {Edge}
  {Devices}.
\newblock Proc. of the Big-{DAMA}, pages 27--32, 2018.

\end{thebibliography}

\section*{Biographies}

\begin{IEEEbiography}[{\includegraphics[width=0.8in, keepaspectratio]{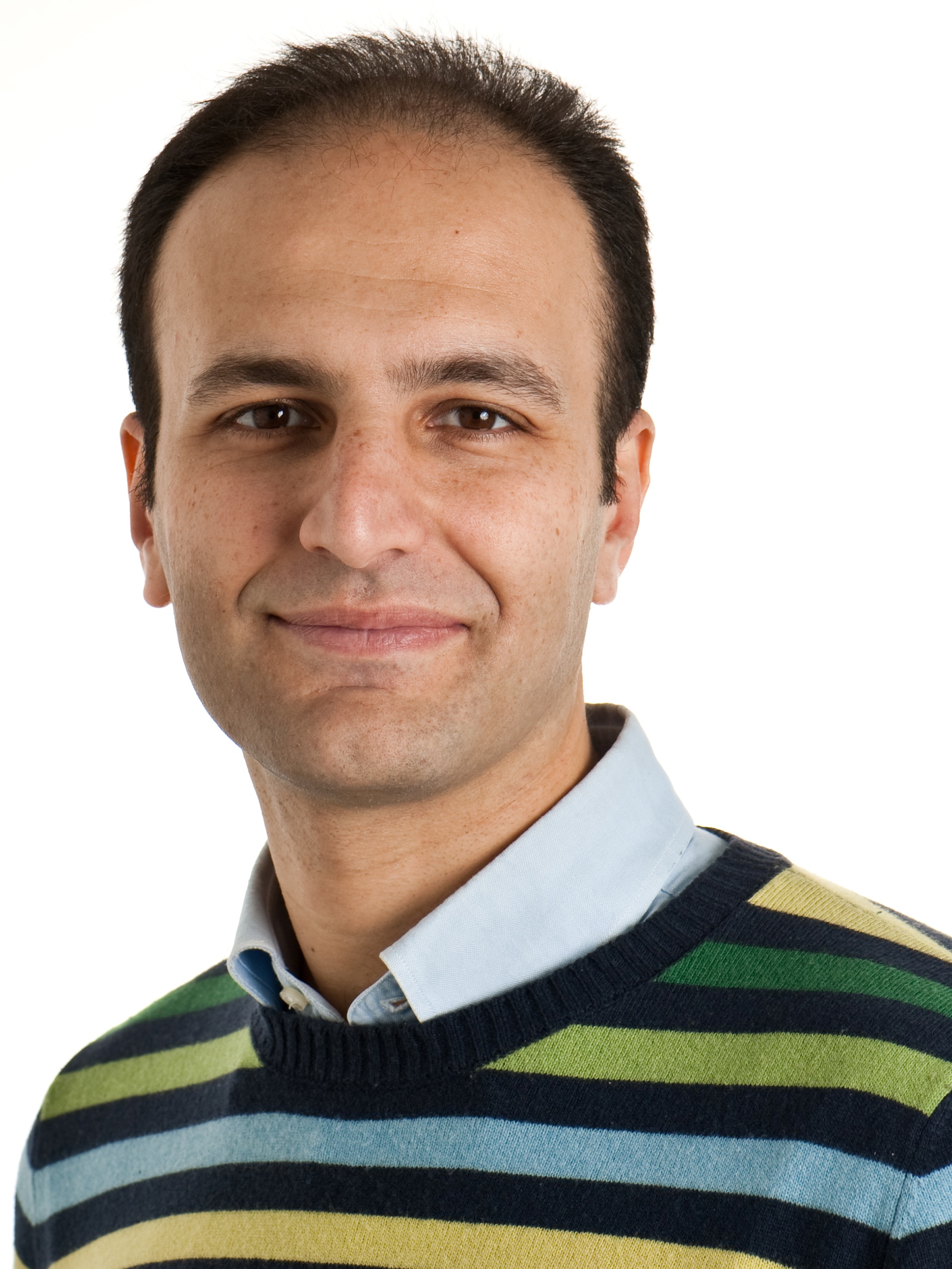}}]{Alessandro D'Alconzo} received the M.Sc. degree in Electronic Engineering with honors in 2003, and the Ph.D. in Information and Telecommunication Engineering in 2007, from Polytechnic of Bari, Italy. Since March 2018, he is head of the Data Science office at the Digital Enterprise Division of Siemens Austria. Between 2016 and 2018, he was Scientist at the Center for Digital Safety \& Security of AIT, Austrian Institute of Technology. From 2007 to 2015, he was Senior Researcher in the Communication Networks Area of the Telecommunications Research Center Vienna (FTW).
%He has been Secretary, and Management Committee representative for Austria, for the COST Action IC0703 ``Traffic Monitoring and Analysis''. He has extensive experience in contributing and managing EU funded projects, as well as in applied research projects in the field of network traffic measurements in collaboration with national telecommunication operators.
His research interests embrace Big Data processing systems, network measurements and traffic monitoring ranging from design and implementation of statistical based anomaly detection algorithms, to Quality of Experience evaluation, and application of secure multiparty computation techniques to cross-domain network monitoring and troubleshooting.
\end{IEEEbiography}

\begin{IEEEbiography}[{\includegraphics[width=1in,height=1.25in,clip,keepaspectratio]{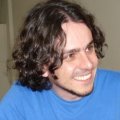}}]%
{Idilio Drago}
is an Assistant Professor (RTDa) at the Politecnico di Torino, Italy, in the Department of Electronics and Telecommunications. His research interests include Internet measurements, Big Data analysis, and network security. Drago has a PhD in computer science from the University of Twente. He was awarded an Applied Networking Research Prize in 2013 by the IETF/IRTF for his work on cloud storage traffic analysis.
\end{IEEEbiography}

\begin{IEEEbiography}[{\includegraphics[width=0.8in,keepaspectratio]{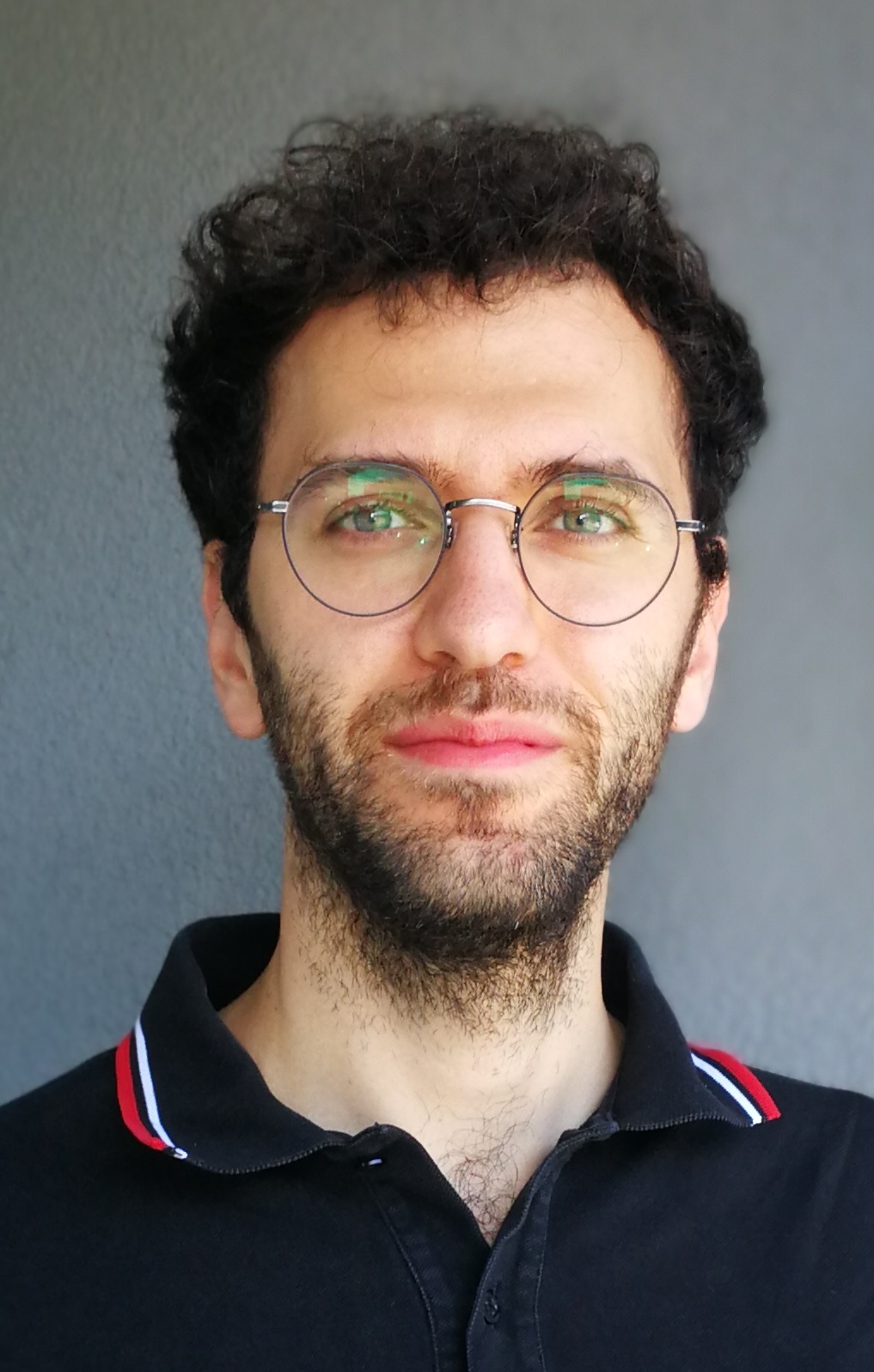}}]{Andrea Morichetta} (S'17) received the M.Sc. degree in Computer Engineering in 2015, from Politecnico di Torino. He joined the Telecommunication Networks Group in 2016 as a PhD student under the supervision of Prof. Marco Mellia, funded by the BIG-DAMA project. In summer 2017 he had a summer internship at Cisco in San Jose, CA. In 2019 he spent six months at the Digital Insight Lab of the AIT Austrian Institute of Technology as visiting researcher. His research interests are in the fields of traffic analysis, security and data analysis.
\end{IEEEbiography}

\begin{IEEEbiography}[{\includegraphics[width=0.8in, keepaspectratio]{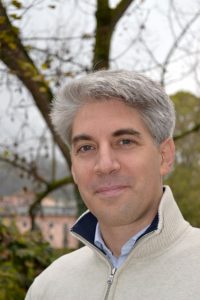}}]{Marco Mellia} (M'97-SM'08) graduated from the Politecnico di Torino with Ph.D. in Electronics and Telecommunications Engineering in 2001, where he held a position as Full Professor. In 2002 he visited the Sprint Advanced Technology Laboratories, CA. In 2011, 2012, 2013 he collaborated with Narus Inc, CA, working on traffic monitoring and cyber-security system design.
% Since 2015 he is collaborating with Cisco Systems for the design of cloud monitoring platforms.
His research interests are in traffic monitoring and analysis, and in applications of Big Data and machine learning techniques for traffic analysis, with applications to Cybersecurity and network monitoring. He has co-authored over 250 papers and holds 9 patents. He was awarded the IRTF Applied Networking Research Prize in 2013, and several best paper awards. He is Area Editor of ACM CCR, and part of the Editorial Board of IEEE/ACM Transactions on Networking.
% He was program chair of several conferences, including ACM CoNEXT'17, IEEE TMA'16 and ITC'15.
\end{IEEEbiography}

\begin{IEEEbiography}[{\includegraphics[width=0.8in, keepaspectratio]{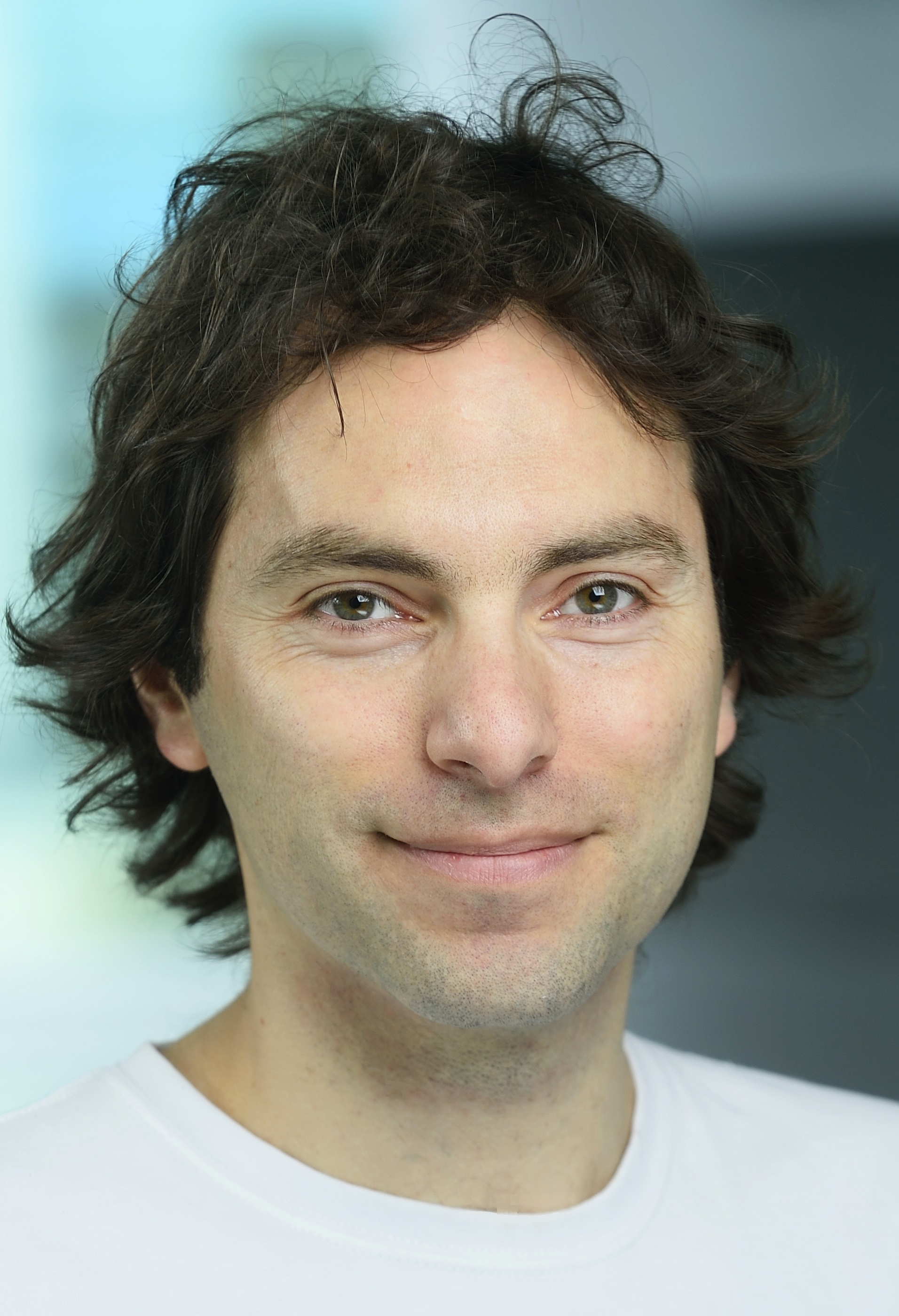}}]{Pedro Casas} is Scientist in AI/ML for Networking at the Digital Insight Lab of the Austrian Institute of Technology in Vienna. He received an Electrical Engineering degree from Universidad de la Rep\'ublica, Uruguay in 2005, and a Ph.D. degree in Computer Science from Institut Mines-T\'el\'ecom, T\'el\'ecom Bretagne in 2010. He was Postdoctoral Research at the LAAS-CNRS in Toulouse from 2010 to 2011, and Senior Researcher at the Telecommunications Research Center Vienna (FTW) from 2011 to 2015.
% He works as technical work leader and project manager for different networking-related initiatives, including research projects and commercial solutions.
His work focuses on machine-learning and data mining based approaches for Networking, big data analytics and platforms, Internet network measurements, network security and anomaly detection, as well as QoE modeling, assessment and monitoring. He has published more than 150 Networking research papers in major international conferences and journals, received 13 awards for his work - including 7 best paper awards. He is general chair for different conferences,
% workshops and leading actions in network measurement and analysis,
including the IEEE ComSoc ITC Special Interest Group on Network Measurements and Analytics.
\end{IEEEbiography}

\end{document}